\documentstyle[12pt,aasms4,flushrt]{article}

\begin{document}

\title{Deep Wide-Field Spectrophotometry of the Open Cluster M67}

\author{Xiaohui Fan}
\affil{Princeton University Observatory, Princeton, NJ, 08544 \\
and
Beijing Astronomical Observatory, Chinese Academy of Sciences, \\
Beijing, 100080, P.R.China\\
Email : fan@astro.princeton.edu}

\author{David Burstein}
\affil{ Department of Physics and Astronomy, Arizona State University, 
Tempe, AZ 85287-1504\\
Email : burstein@samuri.la.asu.edu}

\author{Jian-sheng Chen, Jin Zhu, Zhaoji Jiang, Hong Wu,\\
 Haojing Yan, Zhongyuan Zheng, Xu Zhou}
\affil{Beijing Astronomical Observatory, Chinese Academy of Sciences,\\
 Beijing, 100080, P.R. China\\
Email : chen,zj,jiang,wu,yhj,zzy,zhouxu@qso.bao.ac.cn}

\author{Li-Zhi Fang}
\affil{Department of Physics, University of Arizona, Tucson, AZ 85721\\
Email : fanglz@time.physics.arizona.edu}

\author{Fuzhen Chen, Zugan Deng, Yaoquan Chu}
\affil{University of Science and Technology of China, Beijing, 100039, 
P.R. China}

\author{J. J. Hester, Rogier A. Windhorst, Yong Li}
\affil{ Department of Physics and Astronomy, Arizona State University,
Tempe, AZ 85287-1504\\
Email : jjh@wfpc3.la.asu.edu,raw@cosmos.la.asu.edu,li@samuri.la.asu.edu}

\author{Phillip Lu}
\affil{Department of Physics and Astronomy, Western Connecticut State 
University,\\
Danbury, CT 06810\\
Email : lu@wcsu.ctstateu.edu}

\author{Wei-Hsin Sun, Wen-Ping Chen, Wean-Shun Tsay, Tzi-Hong Chiueh,\\
Ting-Chang Lin, Hui-Jean Guo, Yong-Ik Byun}
\affil{Institute of Astronomy, National Central University,
Chung-Li, Taiwan\\
Email : sun,wchen,tsay,chiueh,lin,batcmgr,byun@phyast.phy.ncu.edu.tw}

\begin{abstract}

We present nine color CCD intermediate-band spectrophotometry of a two square
degree field centered on the old open cluster M67, from 3890$\rm \AA$ to
nearly 1$\mu$.  These observations are taken as a part of the BATC
(Beijing-Arizona-Taipei-Connecticut) Color Survey of the Sky, for both
scientific and calibration reasons.  The BATC program uses a dedicated 60/90 cm
Schmidt telescope with a 2048 $\times$ 2048 CCD and 15 specially-designed
intermediate-band filters to be applied to both galactic and extragalactic
studies.

With these data we show that the BATC survey can reach its goal of obtaining
spectrophotometry to a zero point accuracy of 0.01 mag, and down to V =  21
with 0.3 mag random error. Nine-band spectrophotometry of 6558 stars is
presented. Systematic issues studied include the effect of image undersampling,
astrometric accuracy and transformation from BATC photometric system to 
broad-band systems.

We fit the color-magnitude diagrams (CMDs) with Worthey's theoretical models.
The net result is the excellent fit of the 4.0 Gyr, [Fe/H] = $-0.10$ model to
our data, including a good fit to the main sequence (MS) turn-off.  This fit
predicts $\rm (m-M)_0 = 9.47 \pm 0.16$ and E(B--V) between 0.015 and 0.052.
We show that 16\% of stars in M67 are binaries with mass-ratio larger than
0.7.  Our data are consistent with a toy model with 50\% of the stars
in M67 being binaries and a random distribution of binary mass-ratios, 
although other models with different mass-ratio distributions cannot be ruled
out.

The spatial distribution and mass function (MF) of stars in M67 show marked
effects of dynamical evolution and evaporation of stars from the cluster. Blue
stragglers and binary stars are the most condensed within the cluster, with
degree of condensation depending on mass. The inner part of M67 is missing
most of the lower mass MS stars.  We find M67 to have an elongated shape,
oriented at an angle of $15^{\circ}$ relative to the galactic plane.

Within its tidal radius, the observed MF of M67 between 1.2 $\rm M_\odot$ and
$\rm 0.8 M_\odot$ has a Salpeter slope $\rm \eta = -1.93 \pm 0.66$.  For stars
of mass below 0.8 $\rm M_\odot$, $\rm \eta \sim 0$. It is plausible that the
leveling-off of the MF at lower masses is a result of evaporation of lower
mass stars in this mass range at a rate of one every $\sim 10^7$ years.  
If so, it is plausible that the IMF of M67 has the canonical field value 
of $\rm \eta = -2.0$.  

Overall, we find the stellar distribution as a function of mass within
M67, and the observed MF, are in good agreement with theoretical predictions of
dynamical evolution and evaporation of an old galactic cluster.  Moreover,
the fraction of binary stars and inferred IMF for higher mass main sequence
stars for this old galactic cluster are consistent with known field star values.
This implies a similarity of IMF that persists for at least 4 Gyr in the disk
of our Galaxy.


\end{abstract}

\section{Introduction}

M67 is one of the most-studied old open clusters, because it is reasonably
close, populous ($\sim 1000$ stars) and of similar age as the Sun (cf. Janes
\& Phelps 1994).  As such, photometry of its stars ranges from photographic
(e.g., Johnson \& Sandage  1955, Racine  1971) to CCD (e.g., Gilliand et al.
1991, Montgomery, Marschall \& Janes 1993; hereafter MMJ), and from UBVRI
(Gilliand et al.; MMJ) to uvby (Nissen, Twarog \& Crawford 1987) to DDO (Janes
\& Smith 1984).  The goals of the photometry have been wide ranging.  They
include establishment of photometric standard stars in the cluster (e.g.,
Schild 1983, Chevalier \& Ilovaisky 1991); determination of cluster age and
metallicity (e.g., Gilliland et al.; MMJ; Burstein, Faber \& Gonzalez 1986);
and determination of the physical properties of the cluster, such as
luminosity function (LF), spatial distribution, binary fraction, blue
stragglers, and dynamical state of the cluster, etc. (cf. Racine 1971; Francic
1989; MMJ).

In this paper we report a new photometric investigation of M67 that studies
this cluster over out to a large radius, $\sim 1^\circ$, and down to faint
apparent magnitudes (equivalent of V $\sim$ 20).  The photometric system we
use is new, part of a 15-filter intermediate-band system designed to cover the
wavelength range $\rm 3300 \AA - 1 \mu$, which avoids known bright or variable
sky lines (cf. Thuan \& Gunn  1976), and is equally applicable to galactic
and extragalactic studies.

The M67 observations reported here were obtained both for scientific reasons
and as part of the calibrations done for the Beijing-Arizona-Taipei-Connecticut
(BATC) Color Survey of the Sky.  The BATC survey is a cooperative long term 
program by the institutions represented by the authors of this paper.
The observational goal of the BATC survey is to obtain accurate ($\approx 1\%$)
spectrophotometry for all stellar and diffuse objects in 500 one deg$^2$ 
areas of the sky centered on nearby spiral galaxies, active galaxies, QSO's 
and various calibration fields for Galactic and extragalactic objects, as well
as random fields at high galactic latitudes.
M67 was chosen as one of our first targets specifically because it has been 
well-studied in the past, and because its large angular size provides data 
for a number of tests of the BATC Survey CCD+filter+telescope system.

The observations and reduction of the M67 data are described in \S 2. Special
aspects of how BATC data are obtained and reduced are highlighted. In \S 3,
the combination of large field of view with accurate intermediate-band
photometry permit us to comprehensively investigate both the photometric and
the astrometric accuracy we can attain with BATC survey data. Various
color-magnitude diagrams we can form with these data are presented in \S 4,
where they are used in conjunction with theoretical models to derive estimates
of the cluster age, metallicity, reddening and distance modulus. The binary
fraction of stars on the MS, and the relationship of binaries to intrinsic
scatter on the MS are discussed in \S 5. In \S 6 we discuss the spatial
distribution of stars within M67 as a function of mass, as well as the overall
shape of the cluster.  The luminosity function (LF) and MF of M67 are
discussed in \S 7 in the context of predictions from dynamical evolution of
stars within a cluster that is over 100 relaxation times old.  Our main
conclusions regarding calibration of the BATC survey in general, and our
scientific results on M67 in particular, are summarized in \S 8.

\section{Observations \& Data Reduction}

\subsection{Observations}

BATC observations are done with the 0.6/0.9m f/3 Schmidt Telescope of the
Beijing Astronomical Observatory (BAO), located at the Xinglong station of the
BAO, altitude 900m, located about 150 km NE of Beijing.  A full description of
the BAO Schmidt, CCD and data-taking system, and definition of the BATC filter
system are detailed elsewhere (Chen et al. 1996).  We summarize those details
in this section to the extent needed to define our M67 observations.

The BAO Schmidt telescope is equipped with a Ford 2048 $\times$ 2048 CCD at its
main focus, subtending 0.95 deg$^2$ of sky at 1.67$''$/pixel.  The passbands
of the 15 filters in our system are shown in Figure~1,  atop a typical southern
Arizona night sky spectrum (the Xinglong night sky lacks the sodium-lamp
emission lines).  The central wavelengths and passband widths of the 15 filters
are given in Table~1 along with identifying which filters are used for the
present investigation.  Note that all filters were designed to exclude most of
the bright and variable night-sky emission lines, including the OH forest.

The primary calibration standard stars for the observations are chosen from
flux standard list of Oke \& Gunn (1983). Thus magnitudes in all 15 filters
are on the Oke \& Gunn's AB magnitude system. Transformation from the AB
magnitude system to UBVRI and other photometric systems can be found in Kent
(1985), Edvardsson \& Bell (1989) and Windhorst et al. (1991). We discuss this
problem in detail using M67 observations in \S 3.2. Secondary flux standards
will be developed in a number of ways during the BATC survey, of which
selected stars in M67 from the present investigation form one subset.  Our
goal is to obtain five images of different integration times per filter per
survey field, both in order to increase the dynamic range of our observations
and to search for variable stars and active galaxies.  We wish to reach the
equivalence of V = 21 mag at a signal-to-noise (S/N) of 3 or better in most
filters.  The present M67 observations test the feasibility of this goal.

M67 was observed with the BAO Schmidt and nine BATC filters over six nights in
Jan--Feb 1994.  The log of observations is given in Table~2.  We used the
Number~1 2048 $\times$ 2048 Ford CCD, coated for UV response, with a nominal
quantum efficiency of $\approx 20\%$ at 3200$\rm \AA$, peaking around 44\% at
6500$\rm \AA$, and declining to 6\% at 1$\mu$ (cf. Chen et al. 1996). 
Unfortunately, during the period of observation the UV efficiency of the
coating declined somewhat while also increasing the non-uniformity of the UV
flat-field. Owing to various calibration constraints, we limited the number of
BATC filters observed on a given night to be no more than five.

The No.~1 CCD had a readout noise of 8 e$^-$/pixel with gain adjusted to 3.2
$e^-$/ADU at the time of observation.  When mounted at the f/3 main focus of
the BAO Schmidt telescope, the CCD covers a field of view 58$'$ $\times$
58$'$, and has a linear scale of 1.67$''$/pixel.  The pixel scale somewhat
undersamples the typical seeing conditions ($\sim$ 2-3$''$) of the Xinglong
site.  Thus, one of the tasks of the M67 observations was to test the effect
that this undersampling has on the eventual accuracy of the BATC survey.

The filters were divided into two groups for the M67 observations. Group A has
four filters: 5795$\rm \AA$, 6660$\rm \AA$, 8020$\rm \AA$ and 9745$\rm \AA$,
while Group B has five filters: 3890$\rm \AA$, 6075$\rm \AA$, 7215$\rm \AA$,
8480$\rm \AA$ and 9190$\rm \AA$.  Most of the observations were centered on M67
itself.  For each filter group, one night was devoted to taking short exposure
images offset from the central field by 10$'$ -- 15$'$ in both RA and DEC
direction and pointing the telescope to each of the four quadrants (I, II,
III and IV) relative to the central M67 field (cf. Table~2).  These
``shift-field'' pointings of the telescope permit a series of photometric and
astrometric tests as summarized \S 3.  A total area of 1.92 deg$^2$, centered
on M67, is observed in this way.

\subsection{Flat-Fielding a Degree-Sized CCD Field}

The large field of view of a CCD in a Schmidt Telescope, combined with the
desire to observe many filters in a given night, presents a challenge for
accurately flat-fielding the CCD.  Over a one degree field (twice the angular
size of the full Moon), neither the sky nor the twilight is uniform!
Moreover, our use of intermediate-band filters that avoid bright night sky
lines, combined with our desire to observe five filters per night, means that
we cannot generally get sufficient sky photons for the calibration of the
large-scale non-uniformities of the CCD.  However, to reach our goal of
$\approx 1\%$ spectrophotometry, these large-scale non-uniformities must be
calibrated to an accuracy of $<<$1\%.

After much experimentation (as described in Chen et al. 1996  and Yan et al.
1996), we found that a UV-transmitting diffuser plate, placed in front of the
Schmidt entrance aperture, combined with a standard white-spot dome-flat
system, is quite satisfactory.  The scientific rationale for using such a
diffuser plate is that, for a closed-tube structure such as a Schmidt
telescope, an isotropic illumination at the entrance pupil is physically
equivalent to a homogeneous source of illumination at infinity.  The diffuser
plate provides that isotropic illumination, even when the light source itself
is not isotropic (tests were made even using the full Moon!).

Our tests, detailed in Chen et al. (1996), show that use of the diffuser plate
yields large-scale and small-scale flat-field calibrations that are as good,
if not better, than we can do using the night sky. Since the typical width of
BATC filter is 300$\rm \AA$, the color effect of flat-fielding due to the
color difference of night sky and dome illumination is substantially reduced.
As these flat-field calibrations are taken during the daytime, eight
dome/diffuser flats can be taken with each filter. This means that, combined
with the large full-well capacity of the CCD, our flat-fields are of
signal-to-noise (S/N) $>$700 over the entire CCD.

In addition to the dome/diffuser flat fields, eight bias exposures are taken 
before and after each night.  Dark exposures are not taken, as the dark 
current of the No. 1 CCD is very small (Chen et al.).  All of the data for
a given night:  bias frames, dome/diffuser flat frames (serving the dual 
purpose of simultaneously calibrating pixel-to-pixel and large-scale gradient
variations in the CCD), and program images are then fed into what is called
the ``Pipeline I'' system, written by Zhaoji Jiang for the BATC survey.  

Pipeline I programs subtract the clipped mean of 16 bias frames from each
program frame, and divide it by the clipped mean of the eight flat-field
exposures in the appropriate filter.  The astrometric plate solution is
obtained by {\it a priori} knowing the approximate plate center position, and
then using this information to register the brighter stars in each image with
the Guide Star Catalog (GSC) coordinate system (Jenkner et al, 1990).
Centroids of all the GSC stars in an image, typically $>$100 stars, are used
to determine the plate solution.  The astrometric error of GSC catalog is
claimed to be $0.2''-0.8''$ per coordinate axis.  The observed astrometric
precision is estimated in \S 2.5 using the shift-field observations of M67.
Pipeline I programs also calculate the altitude and azimuth of each image, as
well as the relative direction and phase of the Moon.  An example of the output
of Pipeline I is shown in Figure~2, using one of the M67 centered observations
with the 6075\AA \hspace{1mm} filter.

\subsection{Determination of the Point Spread Function}

The point-spread-function (PSF) of the Schmidt-CCD-filter system was determined
for each observation, using the six analytic PSF forms provided by DAOPHOT II
(Stetson, 1987, 1992).  After some experimentation, the ``Penny'' function was
chosen, as it gave the smallest residual when fitting the data.  The size of
the PSF was permitted to vary linearly across the CCD, owing to an unavoidable
slight tilt of the CCD surface relative to the focal plane. The FWHM of PSF
typically changes $0.1''$ to $0.2''$ across the chip under the conditions when
these observations were made.

Large numbers ($\sim$ 50) of relatively isolated and bright stars were chosen 
to permit study of cross-chip PSF variance.  Once an approximate PSF width is 
obtained, all detected stars near candidate PSF stars are eliminated from the 
original frame and a more accurate PSF is then calculated.  This process is 
iterated several times, first assuming no PSF variation over the CCD image.  
The linear spatial variation of the PSF is then determined after faint stellar 
neighbors of PSF stars are successfully eliminated. 

After a satisfactory model of the PSF and its spatial variation over the CCD
field is obtained, the ALLSTAR algorithm of DAOPHOT II is used to derive the
stellar instrumental magnitudes from the image.  An accuracy estimate is made,
including contributions from readout noise, photon noise and non-perfect 
fitting errors. 

An average FWHM value for each night of observation is given in Table~2, as
measured from the PSF fitting results of filter 6075\AA \hspace{0.3mm} and
5795\AA \hspace{0.3mm} respectively. The limiting magnitudes for each filter
given in Table~2 are estimated from the position where, in a histogram of star
counts vs. apparent magnitude, star counts fall to 1/e the value of the peak.
The limiting magnitude given here refers to the central region of M67 field
(about 20$'$ $\times$ 20$'$) which all six pointings covered. The limiting
magnitude outside this central region, where only one short exposure was
taken per position, is 1.5 mag shallower than in the center.  A more accurate 
determination of limiting magnitude is complicated (Harris 1990) and 
unnecessary for the present paper.
 
\subsection{Calibration to a Standard BATC System}

Each filter group was observed during one photometric night:  17Jan94 for
Group A, 2Feb94 for Group B.  The Oke-Gunn primary flux standard stars
HD~19445, HD~84937, BD+26$^\circ$ 2606 and BD+17$^\circ$ 4708 were observed
during these two nights.  Fluxes as observed through the BATC filters for the
Oke-Gunn stars are predicted by convolving the SEDs of the Oke-Gunn stars with
the measured BATC filter transmission functions.  The night of 17Jan94 had
high winds and poor seeing (cf. Table~2), resulting in somewhat poorer
photometric accuracy for the Group A filters, compared to what is obtained for
 Group B. A summary of the photometric observations is given in Table~3.

BATC magnitudes are defined on the Oke-Gunn system as AB$_{\nu}$ magnitudes
with zero point defined by :

\begin{equation}
\rm m_{i} = - 2.5 log f_{\nu_{i}} - 48.60 .
\end{equation}

\noindent Observations of the standard stars were done with a subset of the CCD
for observing and data reduction efficiency.  This was done after we had 
gained a large degree of confidence in our ability to accurately calibrate 
large-scale sensitivity variations in the CCD (cf. \S 2.2).

Standard stars were taken throughout both photometric nights at a range of air
masses.  In addition, we also used the fact that M67 itself has much UBVRI
data to use a subset of the M67 stars as tertiary standards. Observations of
the Oke-Gunn standards were used to determine the nightly mean extinction
coefficients for each filter.  Both the Oke-Gunn standards and the tertiary
M67 standard stars were used to detect any change in the zero point of
extinction over the night, as such variations are known to exist 
(cf. Colless et al. 1993):

\begin{equation}
\rm m =  a +  m_{0}  + b \times Air\   Mass + f(UT) .
\end{equation}

\noindent No color term is used for atmospheric extinction, as the BATC 
filters are of intermediate ($\rm 300 \AA$) band-width and are
designed to be fairly rectangular in shape.  Verifiable UT 
variations of amplitude $\pm 0.015$ mag were found during both photometric 
nights and were removed from the data.

The initial calibration of the Oke-Gunn stars showed a slight inconsistency 
among the calibration zero points, most likely owing to the known errors
($\sim$ 3--5\%) in determining the filter transmission functions.  Fortunately,
all of the inconsistencies are at the $\sim$0.01 mag level, and can be 
straighforwardly corrected (cf. Yan et al. 1996).

The mean values for the coefficients `a' and `b' in Equation~2, along with
their $1 \sigma$ errors and the net $1 \sigma$ residuals of fitting this
equation to the data, are given in Table~3.  As can be seen, the
night of 17Jan94 (cf. above) has a zero point in the value of `b' about 0.05 
mag higher than for the night of 2Feb94.  This likely indicates  
month-to-month variations in the extinction at the the relatively low altitude 
of the Xinglong site (900m).  Note, however, that despite this change, the net
fitting residual to the standard star observations are quite similar on the
two nights.  The formal error we obtain for the average of standard star
residuals for nine BATC filters is 0.013 mag which, with reasonable estimates
of external errors, indicates that we can define the standard BATC system to
an accuracy of $<$0.02 mag.

\subsection{The BATC M67 Spectrophotometry Catalog}

The calibration of the M67 observations on the photometric nights was applied
to the non-photometric nights in the following manner:

1. There are up to six exposures per filter.  All stars found by DAOPHOT II
in all exposures with photon noise less than 0.02 mag are used to
determine the zero point difference between exposures.  The individual
exposure data for each star are then co-added by weight according to its
S/N.  This gives us one estimate of the S/N of the averaged stellar
magnitudes.  The photometric precision of each star is estimated
by taking into account the noise contribution from every single exposure.

2. The observed magnitudes are then calibrated to BATC standard system by
Equation~2.

3. BATC AB$_{\nu}$ magnitudes and estimated magnitude errors are given for all
filters of each star which is detected in at least 3 filters with magnitude
errors of less than 0.3 mag.  J2000 positions as calculated by the Pipeline I
system (\S 2.2) are given, as well as the Sanders (1977) number for those
stars in M67 so identified.  The catalog that contains these data for 6,558
stars centered on M67 are given in a file which can be electronically accessed
as described at the end of this paper.
 
\section{Accuracy of the Data}

\subsection{Photometric Precision}
 
Sources of CCD photometric error include:  photon statistics of star and sky;
readout noise; random and systematic errors from bias subtraction and
flat-fielding; image defects; possible non-linearity of the CCD; and from the 
PSF fitting and undersampling.  The analysis program DAOPHOT II gives the
theoretical prediction of the error based on the sky and stellar photon
statistics it measures, and tries to estimate the extrapolation error from 
PSF fitting due to the undersampling of the data.

For those stars in common among the five shift-field exposures, we can
estimate the net effect of these sources of error combined.  In Figure~3 we
explicitly calculate the dispersion in their DATPHOT II--measured magnitudes
for all nine filters.  We note again that these shift-field exposures are
relatively short exposures in each filter and, as such, the errors we derive
for faint stars will be lower limits to the precision we can attain with the
BATC survey.

For each filter whose dispersion of measurement is shown in Figure~3, we 
overplot two lines:  The solid line is the observed 1-$\sigma$ error, while the 
dashed line is the theoretical prediction from DAOPHOT II.  The fact that the 
dashed line is difficult to see in these diagrams indicates the high degree of 
consistency with which the BATC Pipeline I + DAOPHOT II programs have 
extracted stellar magnitudes from these images.

From this comparison, we conclude that the photometric precision for a single
short BATC exposure is 0.02 mag for stars brighter than m = 16 in most 
filters, becoming larger for faint stars.  As the BATC survey will have
at least five exposures per filter per field with these exposure times or 
longer, the photometric error for the BATC data is expected to be 0.01 mag for 
brighter stars.  For example, for stars with $\rm m_{7215} < 15.0$ mag, the
observational error from five images is 0.01 mag, increasing to 0.15 mag for
stars with $\rm m_{7215} = 19.5$ mag (equivalent to an effective V mag of 
19.5 -- 20.0) for a single exposure of 3 -- 5 minutes.

Much of the scatter in the shift-field comparison comes from the undersampling
of the PSF.  When the image of the star only occupies a few pixels on the CCD, 
how that image projects on the pixel centers and edges becomes important.  
The difference of sub-pixel position is most serious when a star is near the
limiting magnitude of observation.  An astrometric precision of $\approx
0.3''$, which is what we would expect at the limiting magnitudes of the
shift-field exposures is 0.2 pixels, so that the BATC magnitudes are
sensitive to sub-pixel positioning issues.  Undersampling of the star images
affects the quality of BATC data in two ways: increasing the error of
observation and complicating any calculation of the completeness function (cf.
Stetson \& Harris 1988).

The effect of undersampling on photometric errors can be tested using Group B
filters, as stars in the 3890$\rm \AA$ filter are slightly out-of-focus when
stars in the other four filters are in focus (owing to a difference of 100$\mu$
of optical thickness in an f/3 beam).  The FWHM of the PSF has a mean of
3.4$''$ for the 3890$\rm \AA$ shift-field exposures, which critically samples
(at 2 pixels wide) the PSF. The observed 1-$\sigma$ error is 0.015 mag
compared to 0.013 mag as predicted by DAOPHOT II for bright stars with
$\rm m_{3890} < 13.0 $. In contrast, the FWHM of the
PSF's for the other four Group B filters averages as 2.4$''$, or less than 1.5
pixels, and hence the star images in these filters are undersampled.  In
contrast, the observed 1-$\sigma$ errors for the other four filters is 0.024
mag, compared to 0.027 mag as predicted by DAOPHOT II.

Hence, from this test we estimate the effect of undersampling on our
photometric accuracy to be close to 0.02 mag, as it adds quadradictly to other
random error components. This error accounts for most of the scatter observed
in the photometric comparison for bright stars.  The absolute zero point of
the BATC system is calibrated here using the spectral energy distribution
(SED) of HD19445 (Oke \& Gunn, 1983), to an conservative accuracy of 0.02 mag.

Using the sub-pixel position, we separately examined how the limiting
magnitude of stars depends on the placement of star images relative to pixel
sides and corners.  We find that when the star center falls directly on the
pixel center, $\rm m_{\rm lim}$ = 19.2 for a single 300s exposure in the
7215$\rm \AA$ filter.  In contrast, $\rm m_{\rm lim}$ = 18.7 when the star
center falls on a pixel corner, 0.5 mag fainter.  In addition, while the
histogram of star counts versus magnitude has a steep fall-off at the
magnitude limit for centered stars, it has a much more gradual fall-off for
stars on the corners.  Fortunately, our ability to accurately determine star
centroids relative to pixel placements permits us to better determine
completeness limits, albeit in a complicated manner.

\subsection{Comparsion of the BATC Magnitude System with the UBVRI System}

Of the available photometry of M67 stars, those of MMJ and of Gilliland et
al.(1991) together cover the UBVRI system.  In Figure 4 we present the
relationships that can be formed between the BATC intermediate-band system and
the UBVRI broad-band system.  Stars in common between our data and those of
these two data sources were matched by Sanders (1977) number, generally with
$\rm V < 15.5$.  As such, random errors of observation are minimum for the
comparison of these two data samples to the BATC photometry.

Equations 3--6 show the coefficients of the fits and their r.m.s. errors for
the comparison of our filter colors with the B--V, V--I colors and V and
I magnitudes of MMJ.  Equations 7--10 show the fit coefficients and their 
errors for the comparison with the Gilliland et al.(1991) data (subscript G).

\small
\begin{eqnarray}
\begin{array}{lllll}
\rm (B-V)_{\rm MMJ} & = & + 0.137 &\rm  + 0.474 \times (m_{3890} - 
m_{5795}) &\rm - 0.036 \times (m_{3890} - m_{5795})^{2}   \\
\pm 0.020 && \pm 0.015  & \pm 0.022 & \pm 0.008  
\end{array}
\end{eqnarray}

\begin{eqnarray}
\begin{array}{llll}
\rm V_{\rm MMJ} - m_{5795} & = & -0.034 &\rm  +0.103 \times (m_{3890} - 
m_{5795})   \\
\pm 0.029 && \pm 0.007 &\pm 0.007 
\end{array}
\end{eqnarray}

\begin{eqnarray}
\begin{array}{lllll}
\rm (V-I)_{\rm MMJ}  & = & +0.343 &\rm +1.090 \times (m_{5795} - m_{8020}) 
& \rm + 0.196 \times (m_{5795} - m_{8020})^2   \\
\pm 0.024 && \pm 0.015 & \pm  0.072 & \pm 0.083  
\end{array}
\end{eqnarray}

\begin{eqnarray}
\begin{array}{llll}
\rm I_{\rm MMJ} - m_{7215} & = & -0.362 &\rm  -0.381 \times 
(m_{6660} - m_{8020})   \\
\pm 0.028 && \pm 0.004 & \pm 0.037 
\end{array}
\end{eqnarray}

\begin{eqnarray}
\begin{array}{lllll}
\rm (B-V)_{\rm G} & = & +0.117 &\rm +0.483 \times (m_{3890} - m_{5795}) 
&\rm - 0.029 \times (m_{3890} -m_{5795})^{2}  \\
\pm 0.019 && \pm 0.015 & \pm 0.019 & \pm 0.005 
\end{array}
\end{eqnarray}

\begin{eqnarray}
\begin{array}{llll}
\rm V_{\rm G} - m_{5795} & = & +0.007 &\rm +0.080 \times 
(m_{3890} - m_{5795})   \\
\pm 0.017 && \pm 0.005 & \pm 0.004 
\end{array}
\end{eqnarray}

\begin{eqnarray}
\begin{array}{llll}
\rm (V-R)_{\rm G} &  = & +0.002 &\rm +1.299 \times (m_{5795} - m_{6660}) \\
\pm 0.023 && \pm 0.012 &\pm  0.04 
\end{array}
\end{eqnarray}

\begin{eqnarray}
\begin{array}{llll}
\rm R_{\rm G} - m_{6075} & = & -0.042 &\rm -0.561 \times 
(m_{6075} - m_{6660}) \\
\pm 0.018 && \pm 0.007 & \pm 0.049 
\end{array}
\end{eqnarray}
\normalsize

Comparison of the B--V fits to MMJ and Gilliland et al.(1991) shows evidence
of a systematic difference of $\approx 0.02$ mag between these two data sets,
similar to what was found by MMJ.   Comparison of the V magnitude
transformation also shows evidence of systematic difference, of somewhat larger
size ($\approx 0.04$ mag), in the sense that the Gilliland et al.(1991) V mag
is fainter than that of MMJ at the same $\rm m_{5795}$ BATC mag.  This is
different than what MMJ found, and it is not clear what is the source(s) of
difference.  Despite the differences in transformation coefficients in B--V
and V mag for Gilliland et al. and for MMJ, the errors in the fits are similar,
$\approx 0.02$ mag.  Errors of similar size are seen in the V--I and I mag
fits for MMJ, and the V--R and R mag fits for Gilliland et al.

We are wary of adopting any of these equations as defining the BATC filter
relationship to broad-band colors, owing to these problems.  As such, we
present these relationships only as the external means of verifying our
internal estimates of accuracy, rather than in terms of defining the slopes
and zero points of these relationships.  We leave that task for future work.

\subsection{Astrometric Precision}

J2000 coordinates for stars are obtained using the CCD coordinate frame
established by the Pipeline I reduction package and the best-fit position of
stars given by DAOPHOT II.  The observed astrometric precision is calculated
from the commonly observed stars in the five shift-field exposures, ensuring
that that the full CCD image is sampled.  This comparison only tests for
differences, random and systematic, between the BATC system and the GSC
system.  The astrometric accuracies for all nine BATC filters used in this
paper were separately determined.  Figure~5 shows the diagnostic plot that
results for the 5759$\rm \AA$ filter.

As is evident, the astrometric precision of the BATC survey relative to GSC
system is $\approx 0.15''$ , or about 0.1$''$ per coordinate axis, for
magnitudes brighter than 17, becoming of lower precision for fainter stars. 
Thus, we have confidence that the BATC survey can predict relative positions
of stars at least as accurately as the GSC itself, and are absolutely
calibrated as well as the GSC system.

\subsection{BATC Secondary Standard Stars in M67 field}

Thirty stars are given by MMJ which appear to be suitable local photometric
standards according to six crtieria.  Of these, all but one have observations
in at least four images in each of the nine BATC filters used (one is too
bright and is overexposed).  As these stars do not appear to be variable, we
also consider them suitable as secondary BATC standards. Their positions, nine
color magnitudes, as well as their corresponding MMJ number and Sanders number
(Sanders 1977) are listed in Table~4.

\section{Color-Magnitude Diagrams} 

\subsection{CMD Morphology}

Cluster membership, as determined from the proper motion study of Girard et al.
(1989), is known for the central region of M67 and for stars with $\rm V < 16$.
As our study is both deeper and of much larger angular extent from the
Girard et al. study, contamination by field stars in our analysis will be
handled statistically.

Figures~6--8 plot three representative color-magnitude diagrams (CMDs) that can
be made from our data:  $\rm (m_{3890} - m_{6075})$ vs. $\rm m_{3890}$
(Figure~6) gives us the cleanest view of the CMD; $\rm (m_{3890} - m_{9190})$
vs. $\rm m_{3890}$ (Figure~7) gives us the widest passband colors for these
data that also goes to faint magnitudes; and $\rm (m_{6075} - m_{7215})$ vs.
$\rm m_{6075}$ (Figure~8) gives us the deepest CMD.  In each figure we present two
graphs.  The top graph is the whole data set, covering the field centered on
M67 with a 40$'$ radius (about the tidal radius of M67, see \S6.1).  The
bottom graph contains only the data in the central 15$'$ of the cluster, which
correspond more closely to the data obtained by previous authors. Stars
with known proper motion membership are given distinct plotting symbols, as
explained in the figure caption.  In Figure~9 we replot the $\rm (m_{3890} -
m_{6075})$ vs. $\rm m_{3890}$ CMD, but this time only using stars with
membership probabilities of 80\% or greater.

The morphology of the MS in these CMDs reveals much about the physical 
properties of this cluster.  

a) A well-defined MS exists to $\rm m_{\lambda} \approx 19 - 20$ mag in
each CMD which covers the whole sample, but is less distinct at fainter
magnitudes in the inner 15$'$ sample.  This difference, which pertains to mass
segregation, is discussed at length below (\S 7).

b) The MS has a sharp blue edge and a soft red edge, with the redward scatter
extending to $\sim$0.75 mag brighter than the blue edge.  Such a morphology of
the MS indicates the presence of a populous binary star population in M67 (cf.
Anthony-Twarog et al. 1990; MMJ). The binaries with nearly equal mass form a
distinct sequence about 0.75 mag above the single star MS which is clearly
seen in Figures 6 and 7.

c) The known blue straggler population in M67 is obvious, as well as an
intrinsic dispersion in color and/or magnitude of stars near the main sequence
turn-off.

d) Two gaps are seen along the MS, indicated by the arrows in the lower graphs
in Figures~6--8.  One gap is near the MS turn-off, the other gap is 1.4
magnitudes below the first gap.

e) These CMDs do not extend much brighter than $\rm m_{\lambda} \approx
12$, as even in the shortest exposures (180s) stars brighter than V = 10 are
saturated.  As such, these data do not measure the giant branch of this 
cluster brighter than the red clump stars at $\rm m_{5795} \approx 10$.  

f) The field stars exhibit a well-defined blue color limit in each CMD to
about m = 17; fainter than this, one again finds more bluer stars.  A similar
blue cutoff to the brighter field stars is seen in broad-band CMDs (cf.
Gilliland et al.; MMJ).  We strongly suspect these faint bluer stars are
either white dwarfs, some of which will belong to M67, and perhaps active
extragalactic objects.  We will explore these question further using
spectroscopic data (Burstein et al. 1996).

All of the morphological features of the CMD of M67 we see here have been
either pointed out by previous authors or are apparent on previously published
CMDs (cf. Gilliland et al.; MMJ).  There are two unique aspects about the
present CMDs.  First, the center wavelength of the 3890$\rm \AA$ filter was a
compromise between the stellar and extragalactic observers in the BATC survey.
It covers the Ca II H,K break in the stellar spectrum, making the magnitude
of that filter very sensitive to changes in effective temperature among A--G
MS stars.  As a result, $\rm m_{3890}$ changes rapidly with MS color, so that
a CMD diagram such as Figure~6 or Figure~9 gives a rather vertical
distribution of the MS, highlighting more clearly its morphology.  This is the
reason that the gaps in the MS more noticeable here than in previously
published CMDs.  Second, our observations determine the spectral energy
distributions (SEDs) of bright stars in the M67 field to an accuracy of 0.02
mag. This permits us to investigate the stars in M67 both in terms of the
morphology of the CMDs they form, and in terms of the morphology of their
relative SEDs.

\subsection{Isochone Fitting}

Uncertainties in the determination of the physical parameters of open clusters
are both observational and theoretical in nature.  On the observational side,
uncertainties exist in estimates of reddening (cf. Taylor 1982; Burstein,
Faber \& Gonzalez 1986), metallicity (Burstein et al. 1986), distance (MMJ),
and scatter observed along the MS (cf. \S5, above). On the theoretical
side, authors using different models have obtained different results from the
same data set.  For example, Janes \& Phelps (1994) show that the predicted
age of M67 from ``modern'' CCD photometry ranges over nearly a factor of two,
from 3.4 to 6 Gyr, more depending on whose model is used than which data set
is fit.

For a new filter system such as the BATC filter set, we face the separate
issue of defining these models in terms of our filter sensitivity functions.
As we want to apply our observations equally to individual stars in our
own Galaxy and to stellar populations in other galaxies, we opt for tying our 
filter system into the stellar evolutionary program of Worthey (1992; 1995), 
which also ties our observations to the Lick absorption-line-strength 
measurements of stars, globular clusters and galaxies (cf. Burstein et al. 
1984; Faber et al. 1985; Gorgas et al. 1993; Worthey et al. 1994; 
Trager et al. 1996).  At our request, Dr. G. Worthey has kindly calculated 
isochrones in our filter system for a range of stellar ages and metallicities,
using the known BATC filter sensitivity curves and his stellar evolutionary 
models based primarily on the isochrones of VandenBerg (1985) (see Worthey 
1994 for details) and the theoretical stellar SEDs of Kurucz (1992).

Figures~10 and 11 plot the Worthey-VandenBerg-Kurucz isochrones for a range of
possible ages and metallicities for M67 over the CMDs formed by $\rm (m_{3890}
- m_{6075})$ vs. $\rm m_{3890}$ (Figure~10) and $\rm (m_{6075} - m_{7215})$ vs.
$\rm m_{6075}$ (Figure~11).  We can narrow down the possible range of age and
metallicity from these models in two ways.

First, one can see that the shape of the theoretical CMD near the MS turn-off
does not match that observed in either Figure~10 or Figure~11 for isochrones
older than 5 Gyr.  Second, ages less than 3.5 Gyr require a more metal-poor
stellar population ([Fe/H] $<$ --0.20) than has been indicated by modern
detailed investigations (cf. Burstein, Faber \& Gonzalez 1986).  This, then,
leaves isochrones with ages between 3.5 and 5.0 Gyr and metallicities [Fe/H]
$> -0.20$ as possible candidates.

Our best fit to the CMDs of M67 using these models derive an age of
4.0$\pm$0.5 Gyr and a metallicity of [Fe/H] = $-0.10 \pm 0.05$.  This fit is
shown in Figures~10 and 11. The best-fit isochrone is given in Table~5. This
fit was made to go through the observed main sequence in the $\rm (m_{3890} -
m_{6075})$ vs. $\rm m_{3890}$ diagram, as it is in this diagram that the MS is
most accurately defined.  Our tests (\S 5) strongly suggest that a large
fraction of binaries exist in M67 and their presence contributes much of the
scatter that is seen along the apparent single-star MS in CMD.  As such, the
best fit isochrone was made to fit {\it not} the the middle of the apparent
single-star main sequence, but rather a parallel sequence 0.02 mag redder
(equal to the 1-$\sigma$ photometric error for bright stars) than the hard
blue edge of the MS.

Previous workers (cf. MMJ) have commented on the fact that theoretical
isochrones did not fit well the morphology of the MS turn-off as seen in
broad-band CMDs for any reasonable choice of age and metallicity.  In
contrast, here we find that the 4.0 Gyr, [Fe/H] = --0.10 isochrone from the
Worthy-VandenBerg-Kurucz model matches well the morphology of the main
sequence turn-off in the $\rm (m_{3890} - m_{6075})$ vs. $\rm m_{3890}$ diagram
(Figure~10), but somewhat less well in the $\rm (m_{6075} - m_{7215})$ vs. $\rm
m_{6075}$ diagram (Figure~11).

As the accuracy of our intermediate band magnitudes is comparable to the
accuracy of the broad band BVI data of MMJ (cf. \S4,5), observational errors
cannot be the sole source of this difference.  We will explore another answer
in \S 5 below, when we discuss the effect of binary stars on the M67 CMDs.

In principle the zero point in color and magnitude of the ischrone should
provide the reddening and distance modulus of the cluster, especially when the
isochrone fits the upper MS morphology as well as it does in Figure~10.  The
formal match of the 4.0 Gyr, [Fe/H] = --0.10 isochrone to the CMD in Figure~10
yields $\rm (m - M)_0 = 9.47 \pm 0.05$ and E(B--V) = 0.052$\pm 0.011$, using
the fiducial reddening law of Scheffler (1982) to find E(3890--6075) = 1.8
E(B--V).  Yet, we note that with this fit to the main sequence, the zero point
of the giant branch is too red, owing to the known problem of the Kurucz
models not fitting the SEDs of cool stars well, especially at $\rm \lambda <
4000 \AA$ (cf. Worthey 1994; Clampitt \& Burstein 1996).

This limitation of the Kurucz models also affects our predicted distance
modulus and reddening using the present data.  The CMDs formed by the redder
BATC filters result in a rather small color spread of the upper main sequence
($\sim 0.1$ mag; cf. Figure~11), making these CMDs much more sensitive to
observational errors than colors formed using the 3890$\rm \AA$ filter.  While
the $\rm m_{3890} - m_X$ colors that can be formed with this filter span over
1 mag in upper MS color, the zero points of Kurucz models in these colors are
likely not accurate to better the equivalent of $\Delta$E(B--V) = $\pm$ 0.02
mag (cf. the analysis of Clampitt \& Burstein 1996).  Instead, we use the
limitations put on the possible reddening of M67 by Burstein et al. (1986),
E(B--V) between 0.015 and 0.052 mag, to place an upper limit on the possible
zero point error in the 3890$\rm \AA$ filter to be 0.16 mag.  In turn, this
predicts an upper limit of the error in the predicted distance modulus of 0.16
mag.

\subsection{Main Sequence Gaps}

Existence of a gap in the distribution of stars near the MS turnoff region is
a common feature of old open cluster like M67.  This gap corresponds to the
rapid change of the stellar structure of the stars as they approach hydrogen
core exhaustion.  The shape and size of the MS turnoff, and the relative
position of the gap, is related to the effect of convection overshooting (cf.
Demarque et al., 1994).  In the isochrones, this gap is usually represented by
a turn to the blue by the MS just below the actual turn-off luminosity.  As
can be seen in Figure~10, this is exactly where the Worthy-VandenBerg-Kurucz
isochrone matches the observed gap in the stars, giving both added weight to
the reliability of this fit and support to this interpretation of the gap.

The existence of a second MS gap, 1.4 mag in the 3890$\rm \AA$ filter below
the first gap, is not {\it a priori} expected.  Whereas the brighter gap is
almost devoid of stars, the fainter gap is simply deficient of stars.  We see
this second gap in all CMDs made from the BATC filters, in all radial
distances from the cluster center, as well as in the broad-band BVI CMDs of
MMJ.

The question raised by this fainter gap is whether it is a statistical fluke
in the initial mass function (IMF) of the cluster, or it is an indication of 
some physical process during stellar evolution that only affects a certain 
range in MS mass.  Although similar gaps were seen in the broad-band CMDs 
published by McClure et al. (1981) for NGC 2420 and NGC 2506, of similar age 
and metallicity as M67, a more recent uvby investigation of NGC 2420 by 
Anthony-Twarog et al. (1990) shows little evidence of this gap.

Interestingly, this second gap appears at about a mass of 1 $\rm M_{\odot}$,
where the calculations of Lattanzio (1984) point out that grain sedimentation
in the forming star could increase the heavy metal content of the stellar
core.  The isochrone then calculated upon this assumption does have a second
gap in the MS near stars of 1 $\rm M_{\odot}$.

This second gap (and the brighter gap) are easy to see in the CMDs of M67 
owing to the large number of stars in this cluster ($\sim 1000$, see below).  
These gaps may not be as evident in CMDs of less populous clusters, or in CMDs
with data of accuracy $>$0.05 mag in magnitude, nor will it be easily
visible in CMDs in which the color coverage of M67 stars is relatively small 
(Fig~8). Future high-precision CCD photometry of other open clusters of 
similar age and mass as M67 should settle the question of whether this second 
gap is a statistical aberration or an indication of an important physical
process in stellar evolution.

\section{Percentage of Binaries among M67 Main Sequence Stars}

The existence of a large binary population among the M67 MS stars is
well-known (cf. Racine 1971; MMJ).  It is evident in the CMDs of Figures~6--11
as the red ``fuzziness'' and blue hard edge of the MS, and as a scatter above
the MS that is primarily within 0.75 magnitude. These facts have influenced
other workers (cf. MMJ) to assume that M67 contains a bimodal distribution of
binary star mass ratios, with some binaries formed of stars of near equal
mass, and others combine stars of very unequal masses.

We use the precision of the present data set to rexamine the mass ratios and
number of binaries in M67 from the following point of view:  what would be the
distribution of stars along the MS if all stars are in binaries, but the ratio
of luminosity of secondary to primary is permitted to randomly vary from zero
to unity?

We can look at the magnitude-color distribution of MS binaries in 
one of two ways: either in terms of a fixed primary mass, but varying secondary
mass (termed iso-primary-mass binaries), or with same secondary/primary mass 
ratios (iso-mass-ratio binaries).  The top graph in Figure~12 plots tracks of 
iso-primary-mass binaries as solid lines, while dashed lines are tracks of 
iso-mass-ratio binaries.  The bottom graph in Figure~12 plots these predictions
over the relevant subset of the actual M67 $(\rm m_{3890} - m_{6075})$ vs.
$\rm m_{3890}$ diagram.  From inspection of the two parts of Figure~12 we can 
make four points:

a) All binaries with mass ratios (secondary mass/primary mass) from 0.0 to 0.5
are within 0.1 mag of the MS, even using a filter as blue as $\rm
\lambda_{\rm eff} = 3890 \AA$.  To photometrically distinguish the low mass
ratio binaries from single stars at the $>3$-$\sigma$ level in CMDs requires
measurements accurate to $\le 0.03$ mag, near the limit for stars in crowded
fields.  Fortunately, the field around M67 is not too crowded, and the proven
accuracy of the BATC magnitudes is 0.02 mag for brighter stars.

b) Binary stars with mass ratios 0.9 to 1.0 aggregate near the equal-mass
binary line.  The fact that there are several stars with high membership
probability signficantly beyond the equal-mass binary line in Figure~12
suggests that some multiple star systems also exist in M67.

c) Binaries with mass ratios between 0.5 and 0.9 are spread out between the
apparent single star MS and the apparent equal-mass binary MS. The above
points show that an accurate CMD will have the appearance of having mostly
equal mass binaries and single stars, rather than binaries with a continuity
of mass ratios.

d) Near the MS turnoff, the luminosity contrast between the primary and
secondary star increases, causing all binaries to have photometric properties
more like single stars.  This is especially true for red colors, where the
color contrast between primary and secondary stars is smaller than for blue
colors.  Thus, at its upper end, the binary MS narrows, leading to a
noticeable ``fuzzy'' appearance to the upper main sequence evident in Figures 
6--9 and in other published CMDs of this cluster (cf. Gilliland et al. 1991;
MMJ).

We further simulate the BATC filter observations using a ``toy'' open cluster
with 1000 stars under the following assumptions:  a) Main sequence stars
follow the best-fitting isochrone of 4 Gyr, [Fe/H] = --0.10. b) The slope of
MF is independent of mass for the MS, as we derive in \S 8 (below) for lower
MS.  c) A given star has a 50\% chance of being a binary. d) The distribution
of mass-ratios among the binary stars is random.  e) The magnitudes of the
stars are measured with a photometric precision of 0.02 mag in each passband.
f) The primary stars have masses between 0.7 $M_{\odot}$ and 1.3 $M_{\odot}$.
The simulated CMDs of $(\rm m_{3890}-m_{6075})$ vs. $\rm m_{3890}$ and $(\rm
m_{6075}-m_{7215})$ vs. $\rm m_{6075}$ are shown in Figure 13. Comparisons of
Figure 13 with Figures 6 and 8 show no qualitative difference in CMD
morphology. The existence of an obvious photometric binary branch in the CMD
of $\rm (m_{3890}-m_{6075})$ vs. $\rm m_{3890}$ as well as the large scatter
near the MS turnoff in both the observed and simulated CMDs, suggest that M67
contains a large fraction of binaries.

We have calculated the star number distribution relative to the MS, taking the
magnitude difference of each star from the fitted main sequence isochrone,
independent of whether or not that star has a proper motion membership
estimate.  In Figure~14 we compare the observed distribution relative to the
prediction of our toy model cluster which has 50\% binaries based on the CMD
of $\rm (m_{3890}-m_{6075})$ vs. $\rm m_{3890}$ for the color range in 
$\rm (m_{3890}-m_{6075})$ between 1.4 and 2.5 (colors are used, as binaries
are most constant in color than in absolute magnitude).  It is evident that 
model and real distributions are quite similar.

A noticeable feature of the real histogram distribution is that it is markedly
asymmetric; more stars scatter to the red of the peak than to the blue. This
asymmetry is also evident in the $\rm (m_{3890}-m_{6075})$ vs. $\rm m_{3890}$
CMD (Figure~6) as a more sharply-defined blue edge to the main sequence than to
the red edge (cf. discussion by MMJ).  The model does very well in fitting this
asymmetry, strongly implying that within what appears to be the single-star
main sequence are also contained a substantial number of low mass-ratio binary
stars.

If we do a straightforward accounting of obvious binaries, we find that 16\% of
the stars in M67 within 33.3 $'$ from the center are binaries with mass ratios
0.7 or higher according their position on the CMD of $\rm (m_{3890}-m_{6075})$
vs. $\rm m_{3890}$.  The background contribution is estimated and subtracted
in the way described in \S 6.1.  If we restrict our sample to the area
observed by MMJ, we would derive an observed binary frequency of 22\%, the
same as derived by MMJ in the same mass-ratio interval.

The true binary fraction in this cluster depends critically on how we
determine the contribution from low mass-ratio binaries.  The asymmetric
blue/red distribution of stars along the main sequence shows they exist and is
consistent with the dynamical evidence discussed in \S 6.1.  A more
quantitative determination of their numbers is difficult both owing to errors
in background determination and especially to the fact that it is difficult to
photometrically distinguish low mass-ratio binaries from single main sequence
stars.  MMJ tried to model the background and got a fraction of 38\% for
binaries of all mass-ratios.

As the agreement between model and data in Figure~14 shows, the most
straightforward interpretation is that the binary mass-ratio distribution is 
random between 0 and 1.  If so, then 50\% of the stars in this cluster are 
binaries.  However, by the same token, exactly how many low mass-ratio binaries
are in this cluster is confused precisely because most of the low-mass ratio
binaries have photometric properties so similar to those of single stars.
As a result, while it is apparent there are many low mass-ratio binaries in
M67, we also cannot rule out the binary fraction being somewhat less than 50\%, 
or somewhat more.

The mass-ratio distribution of field binaries has been widely-studied. Trimble
(1978) pointed out that the apparent distribution of binaries has two peaks in
mass ratio, at 0.3 and 1.0.  Previous investigators who estimated the binary
frequency among cluster MS stars have used such a bimodal distribution for
their estimates (cf. Anthony-Twarog et al. 1989; Mazzei \& Pigatto 1988; MMJ).
Later analyses of different samples of field binaries however, give
very inconsistent results on the mass ratio distribution from bimodel to quite
randomly distributed (cf. Trimble, 1990).  In the case of M67, the binary 
distribution is more consistent with being randomly distributed than being 
double-peaked.

\section{Spatial Distribution of Stars}

\subsection{Mass-dependent Spatial Distribution}

It is well-known that M67 is sufficiently populous and dense to have a
relaxation time much shorter than its age.  For example, Francic (1989) gives
the relaxation time of M67 to be $1.7 \times 10^{7}$ yr, while Mathieu \&
Latham (1986) calculate $10^{8}$ yr.  According to Binney \& Tremaine (1989;
pg. 526), the time it takes a star cluster to evaporate away is $\sim 100$
times the relaxation time, or in the case of M67, 2--10 Gyr (cf. calculations
by Francic). As the estimated evaporation time is comparable to the age of the
cluster, we would expect to see marked effects of mass diffusion, such as mass
segregation among MS stars. With the present data set we can explore these
issues explicitly.

Dynamical evolution of the cluster over time will progressively move the lower
mass objects outward and the higher mass objects inward.  The lowest mass
objects in a cluster are, of course, the lower MS stars.  The higher mass
objects include the upper MS stars, subgiant stars and giant stars, as well as
the binary stars.  Dynamical evolution of the binary stars is likely to be
separate from that of single stars, owing to the fact that energy can be
exchanged between the binary orbits and the cluster orbits (cf. discussion by
Binney \& Tremaine, Chap. 8).

Until the present study, accurate photometric data of M67 stars were limited
spatially, or large-scale studies had bright magnitude limits.  Furthermore,
estimation of field star contamination, important for understanding the low
mass end of the MS of M67, was problematic.  As a result, while MMJ were aware
of the likelihood of mass segregation effects in their estimates of the cluster
LF, they could not explicitly measure it.

The present observations are well-suited in terms of field size and depth
of study.  Field star contamination can be estimated with the present data set 
given a reasonable value for the tidal radius of M67.  According to Von Hoerner
(1957), the tidal field of the Galaxy produces a tidal radius $\rm R_{\rm t}$ 
defined as: 

\begin{equation}
\rm  R_{\rm t} = R_{\rm G} \times (\frac{\rm M_{\rm c}}{\rm 2M_{\rm G}})^{1/3},
\end{equation}

\noindent where $\rm R_{\rm G}$ is distance to the galactic center, 
$\rm M_{\rm G}$ is the total mass of the Galaxy within $\rm R_{\rm G}$.  
Outside of $\rm R_{\rm t}$, the tidal field of the Galaxy is predicted to 
strip stars away from M67, so beyond this radius can be considered the 
``field.''  Francic (1989) gives $\rm R_{\rm t}$ = 42$'$ for M67 (using the 
distance modulus of Francic).  By coincidence, the largest radius from the 
center of M67 at which our data can reasonably sample stars in a full annulus 
is 42.5$'$.  Hence, we choose an annulus of 37.5$'$ to 42.5$'$ from the 
center of M67 to sample the ``field.''  A star outside of the tidal radius is 
also not necessarily being evaporated from the cluster immediately.  So by 
choosing the annulus in this way, we are slightly underestimating the number 
of the very outer-most M67 stars.

Figure~15 shows the apparent magnitude function of the field stars.
The differential star counts A(m) between $m_{5795}$ of 14 and 18 is roughly
proportional to $10^{-0.2m}$. This is consistent with the prediction
based on galactic models (Bahcall \& Soneira, 1980, 1984), indicating that
the contribution from M67 members in this annulus is negligible. 

Figure~16 plots the cumulative radial distributions for seven subsets of the
M67 stars, with their estimated field star contamination removed: 
subgiants+giants, three absolute magnitude ranges of MS stars, blue
stragglers, and two absolute magnitude ranges of obvious photometric binary
stars.  Following Mathieu \& Latham (1986), we give a lower limit of $\rm 2
M_\odot$ to the mass of blue stragglers and assume they should be compared to
the binary distribution, not the single star distribution. Stars are
classified based on their positions in the $\rm (m_{3890} - m_{6075})$ vs.
$\rm m_{3890}$ CMD and the best-fit isochrones given in \S 4.2. The binaries
are those with mass ratios estimated according by the binary model of \S 5
as being 0.7 or larger.

Mathieu \& Latham (1986), Francic (1989) and MMJ all give evidence of mass
segregation in M67.  The mass and half-mass radius for each stellar subset of
the present data are given in Table~6, and span an average mass range from 0.7
$\rm M_\odot$ to at least 2 $\rm M_\odot$.  Lower limits are given for the
masses of MS stars, subgiants and giants, owing to the existence of binaries
with small mass-ratios.

Interestingly, if we separate our sample into mostly single stars plus
low-mass-ratio binaries, versus obvious photometric binary stars and blue
stragglers, a significant trend is found in terms of mass segregation: Among
the higher mass objects, blue stragglers are the most concentrated and thus
implied to be most massive; obvious binaries of lower mass are the least 
concentrated, but are still more concentrated than any of the MS ``single'' 
stars.  Among the ``single'' stars, there is some evidence that degree of
conentration increases with increasing mass, albeit with somewhat large errors.

All of the nominal ``single'' MS stars have half-mass radii signficantly smaller
than all of the binary stars. This pattern is consistent with the suggestion 
that the more massive binary stars dynamically behave almost as if in a 
different cluster than the single stars (cf. Binney \& Tremaine, 1989; 
Chapter 8).  Both numbers and high masses are consistent with the idea that 
blue stragglers evolve from near equal-mass-ratio binary systems 
(cf. Mathieu \& Latham, 1986).

We also investigated if we could detect mass segregation among the low
mass-ratio binary stars whose photometric properties place them near the
single-star MS.  Using the fitted isochrone in the $\rm(
m_{3890} - m_{6075})$ vs. $\rm m_{3890}$ CMD and in the same magnitude level as the other tests on
MS stars, we separated the stars into ``blue'' side (-0.05 to 0.02 mag from
the isochrone) and ``red'' side (0.02 to 0.09 mag from the isochrone) of the MS
(MS `b' and MS `r', respectively in Table~6), and then calculated their
spatial distribution.  While we do see a difference in the expected sense,
the difference is small relative to the observational error.

The data in Table~6 and plotted in Figure~16 provide ample evidence 
that dynamical evolution has substantially modified the distribution of stars
in M67.  Indeed, M67 has a halo made mostly of faint low mass stars and
a core made primarily of high mass binaries plus giant, subgiant and upper main
sequence stars.  It is apparent that, in M67, we are looking at a cluster
in the process of evaporation.  Below, we suggest one way the evaporation
rate might be calculated in the observed mass range.

\subsection{The Shape of M67}

Our data can give a hint of the two-dimensional shape of M67.  In Figure~17 we
plot the CMDs obtained at nine different positions around M67. The inner CMD
comes from a radius of $<5'$ of the cluster center, while the outer 8 CMDs are
taken in pie-shaped cuts of $\rm 45^\circ$ angle, between radii of 5$'$ and
16.67$'$.  These cuts are positioned in Figure~17 such that the north cut
is up and the east cut is to the left. 

A well-defined M67 MS is evident from the NE to SW direction, but is
ill-defined in NW and SE directions.  To better see this, we counted the number
of stars along main sequence locus vs. all other stars as background between
$14 < \rm m_{3890} < 19 $ in the eight cuts around the center. We found 35.5 $\pm$
0.6 MS stars per area in NE, S, SW, W panels and 22.5 $\pm$ 1.7 MS stars per
area in N, NW, SE, and E panels. In contrast, the number of background stars
per region is 21.5 $\pm$ 3.3.  Thus we see a $>3 \sigma$ difference in MS star
counts between the NE-SW direction and the NW-SE direction.

This indicates that M67 is elongated at a position angle of $30^{\circ}\pm
45^{\circ}$, or $15^{\circ}\pm 45^{\circ}$ relative to Galactic plane.  It is
not too surprising that a Galactic cluster as old as M67 has an elliptical
shape, given the tidal stresses to which has been subjected over its lifetime.
Terlevich (1987) used N-body simulations to predict that the flattening of
evolved open clusters will produce major axes parallel to the Galactic plane.
Our result here is consistent with her simulation.

\section{Luminosity Function and Mass Function}

\subsection{Luminosity Function}

The previously-derived LF for M67 show a maximum near the MS turnoff, and a
decline with fainter main sequence magnitudes (van den Bergh \& Sher, 1960,
Francic 1989; MMJ).  MMJ pointed out that this decline could be an artifact of
mass segregation and mass loss.  Certainly the analysis of the previous
section supports this view, which we explicitly examine here.

To accurately derive the faint end of the LF for M67, we need to understand
our selection criteria, which means determining the limiting magnitude of our
data as a function of distance from the cluster center.  As discussed before
(\S 3), our data are comprised of many exposures obtained under different
observing conditions and subject to different magnitude limits, owing to
different exposure times and effects of undersampling.  In particular, the
longest exposures were taken of the cluster center, over a nearly one-degree
field of view, while the extension of our data to twice this angular extent
was done with shorter exposures.

Thus, we decided to limit the LF analysis to the passband that goes the
deepest across the whole region sample --- 7215$\rm \AA$. To avoid
undersampling problems, we limit our analysis to stars which have $\rm
m_{7215} < 18.0$ mag.  The correction for field star contamination is done in
the same way as in the spatial distribution analysis.

Figure~18 shows the LF for M67 as found in three volumes: a) from 0 to 5$'$
(``cluster core''); b) 0 to 16.67$'$ (``intermediate cluster''); and c) 0 to
33.33$'$ (``whole cluster'').  We find that the 7215$\rm \AA$ LF for the
``cluster core'' peaks at $\rm m_{7215} = 12.5$ mag; one magnitude below the
MS turn-off.  This is consistent with what Francic (1989) and MMJ found.
Contributions from binaries are included when computing LF.

In contrast, the LF for the ``intermediate cluster'' rises quickly at the MS
turnoff, and then remains fairly flat within errors to the magnitude limit.
The LF for the ``whole cluster'' reinforces the trend seen in the intermediate
cluster, showing a slight trend of increasing star numbers down to the
magnitude limit.  There is also a marked deficit of stars at $\rm m_{7215}
\approx 16.5$, corresponding to $\rm V \approx 16$, which only appears when
the outer annuli of M67 are included.

Hence, previously voiced suspicions (cf. MMJ) about the effect of mass
segregation on the observed LF of M67 are confirmed. As is even clear from the
CMDs in Figures~6--9, low mass MS stars are predominantly found in the outer
parts of M67, while the higher mass main sequence stars are found in the inner
part.

\subsection{Observed Mass Function and Effects of Dynamical Evolution}

The MF for stars is commonly written in the form of a Salpeter (1955) power
law: $\rm dN(M) \propto CM^{\eta} dM$.  Scalo (1986) finds $\eta = -2.00 \pm
0.18$ for field stars. Francic (1989) finds a similar slope for six young
clusters ($\eta = -1.97 \pm 0.17$).  In contrast, Francic derived $\eta =
+3.63$ for NGC 752, another old open cluster and +2.49 for M67 in the V
passband.  Prata (1971) theoretically studied the evolution of the MF of M67,
and concluded that IMF of M67 is likely to be deficient in low-mass stars. 
With our current sample we can test this idea explicitly.
 
We use the best-fitted VandenBerg (1985) isochrone (i.e., 4 Gyr, [Fe/H] =
--0.10) to establish the relationship between absolute magnitude and mass to
derive the LF for M67.  Field star contamination is corrected as before. We
fit the data with a Salpeter power law within mass range of 0.5 $\rm M_\odot$
to 1.2 $\rm M_\odot$. The results of the fitting are given in Table~7 as a
function of increasing volume sampled in M67.  As is evident, we obtain a MF 
of +1.12 for the cluster core, but --0.51 for the whole cluster.

Figure~19 shows the MF for the whole M67 cluster (to a radius of 33.33$'$), to
a limiting mass of 0.5 $\rm M_\odot$ up to the MS turnoff.  A slope of $\eta =
-2.0$ is shown superimposed on the massive end of the MS. A formal MF for
stars with masses between 1.2 and 0.8 $\rm M_\odot$ is fitted by a power law with
$\eta = -1.93 \pm 0.66$.  It is evident that the slope of the MF changes from
having $\eta \sim -2.0$ for the mass range 1.2 -- 0.8 $\rm M_\odot$ to $\eta
\sim 0$ (i.e., independent of mass) for lower mass stars (0.8 -- 0.5 $\rm
M_{\odot}$).  Any residual contamination by M67 member stars in our star
background is unlikely to signficantly change the lower mass MF slope. 
Similarly, from the excellent fit to the Bacall and Soneira model, especially
at the fainter end (Figure~15), we conclude the change of slope in the MF at
0.8 $\rm M_\odot$ is real.

The fact that we observe the effects of dynamical evolution among the stars
that currently define M67 as a cluster strongly suggests that low mass stars
have selectively evaporated from the cluster over time. This is consistent
with theoretical estimates, which place the evaporation time scale for M67
comparable to its age (cf. Prata 1971; Francic 1989; \S 6.1 above).  One kind
of quantitative estimate of the number of lost stars (at least down to a mass
of 0.5 $\rm M_\odot$), can be made if we assume that M67 started out with a
Scalo IMF slope of $\rm \eta = -2.0$.  We then take this best fit $\rm \eta =
-2.0$ slope for masses 1.2--0.8 $\rm M_\odot$, and extrapolate this to the
lower MS masses.  From this we predict $\sim$450 stars in the mass range 
0.8 to 0.5 $\rm M_\odot$ have evaporated from this cluster over its 4 Gyr age, 
or at an average rate of one star per 10 million years.  Such a mass loss 
rate, while not proven by this calculation, is plausible.  From this 
comparison we conclude that it is equally plausible that the IMF of M67 
followed a Salpeter power law with the canonical field star value of 
$\eta = -2.0$.

Our observed mass estimate for M67 is larger than previous estimates owing to
the fact that we have sampled a larger volume of the cluster to fainter limits
than previous surveys (cf. Mathieu 1985; Francic 1989; MMJ). In Table~7 we
compute the observed mass of M67 within an increasing radial distance from the
cluster center, giving both the single star estimate and that we would
estimate from both binary and single stars. 

\section{Conclusions}

These M67 observations are but the first of many such observations of galactic
clusters and galactic fields pertaining to studies of Galactic structure,
which we will obtain with the BATC survey.  By using this open cluster as a
calibration source, we are able to show the accuracy of the data we can obtain
with the 0.6m/0.9m BAO Schmidt telescope at its Xinglong, China site.  It is
evident we can obtain spectrophotometry from the ultraviolet to $\sim 1\mu$
down to an intrinsic accuracy of better than 0.02 mag for all objects in the
nearly 1 deg$^2$ field of the CCD, using Oke--Gunn primary standard stars (Oke
\& Gunn 1983).  Similarly, we can define the positions of these objects to an
accuracy of 0.15$''$ for bright stars, using the Guide Star Catalog (Jenkner et
al. 1990). To reach these accuracies, and to be able to observe through as
many as five filters in a given night, we have used a diffuser plate in
combination with our dome flat fields to obtain flat fields at least as 
accurate as we could obtain using sky observations.

Our observations of M67 span an area $1.92^\circ \times 1.92^\circ$ centered
on the cluster in nine BATC intermediate-band filters.  The color magnitude
diagrams formed from these data show morphologies consistent with previous CMDs
of M67, as well as defining better than most the gap in the MS near the MS
turn-off.  Isochrones of a range of appropriate stellar populations has
kindly been made available by Dr. G. Worthey.  These models, which convolve
our filter senstivity cures with his theoretical spectral energy distribuitions
(based on the theoretical stellar evolutionary tracks of VandenBerg (1985) and 
the synthetic stellar spectra of Kurucz (1992)), have been fitted to our data.

We show that the combined Worthey-VandenBerg-Kurucz model fits the $\rm
(m_{3890} - m_{6075})$ vs. $\rm m_{3890}$ CMD very well for an age of 4 Gyr and
[Fe/H] = --0.10, yielding a reddening of E(B--V) between 0.015 and 0.052 mag
and a distance modulus $\rm (m-M)_0 = 9.47 \pm 0.16$ mag. The uncertainties in
the derived distance modulus and cluster reddening are dominated by
uncertainties in the predicted Kurucz fluxes below 4000$\rm \AA$.  As such, we
use the range of known reddenings for M67 to place limits on what are these
uncertainties.

It is well known that M67 has a parallel binary main sequence. We find that
16\% of stars in M67 are binaries with mass-ratio larger than 0.7, in agreement
with previous studies. In a departure from previous work, here we model the 
binary star population of M67 in terms of a random distribution of 
secondary/primary mass-ratios from 0.0 to 1.0.  We find that even if the 
mass-ratio distribution of binaries in the cluster is random, there will still 
be an apparent offset, parallel MS in the CMD.  Although we cannot accurately
determine the fraction of low mass-ratio binaries in the cluster, our data are
consistent with 50\% of the observed stars in M67 being binaries.

As our survey of M67 combines deep images, accurate photometry and a wider
field coverage than previous surveys, we are able to explore dynamical
evolutionary issues pertaining to this old galactic cluster.  We find much
evidence of substantial dynamical evolution of M67.  The spatial distribution 
of stars is clearly dependent on their masses.  Blue stragglers are the most 
centrally--condensed among the stars, consistent with previous observations 
(cf. Mathieu \& Latham 1986). The assumption that blue stragglers are binary
stars of nearly equal mass ratio is consistent with the steady decrease in
central concentration observed from blue stragglers to lower mass binaries 
with secondary/primary ratios $>$ 0.7. A similar trend is seen among the
``single'' stars --- giant star to lower mass main sequence stars --- where
the word single is in quotes owing to a likely contamination by low-mass-ratio
binary stars.  Overall, the binary stars as a group are more centrally 
condensed within M67 than are the ``single'' stars.  With these data we can 
also investigate the two-dimensional shape of M67, which we find to be
elongated along an angle of $15^{\circ} \pm 45^{\circ}$ relative to the 
Galactic plane. 

Similarly, we find that the luminosity function (LF) of M67 is dependent on the 
volume samples. If we observe only the inner core of the cluster, we find 
results similar to those previously obtained, in which there are more stars on 
the upper MS than on the lower MS.  However, if we include the cluster stars
out to a radius of 33.33$'$ (within the tidal radius estimated by Francic 
(1989)), we find a LF that rises from the MS turnoff and then flattens out at 
fainter absolute magnitudes.

We use the isochrone fit of the Worthey-VandenBerg-Kurucz model, combined
with our model of binary star distribution, to derive a mass function (MF) for
M67 at its present age.  Using the Salpeter (1955) definition of a MF
power law, we find the slope $\eta$ (as in $\rm M^\eta$) to be near the 
canonical field star value of --2.0 for MS stars with masses betwen 1.2 and
0.8 $\rm M_\odot$, but which levels off for lower mass stars to the limit
of our observations (at 0.5 $\rm M_\odot$).  If we attribute this leveling off
as due to evaporation of stars through dynamical evolution of this old cluster,
we estimate that M67 has lost one star of mass between $\rm 0.8 M_\odot$
and $\rm 0.5 M_\odot$ every $\sim 10^7$ years.

Hence, with these data we have been able to observe both direct and implied
evidence of substantial dynamical evolution of M67, consistent with theoretical
expectations.  We note that our data show an old open cluster whose most
likely binary fraction and IMF are remarkably consistent with that of the field.

Files in standard ADC format, giving the full BATC data for all 6558 stars
can be accessed either via anonymous ftp from samuri.la.asu.edu (IP
129.219.144.156), in the subdirectory pub/m67, or from the Astronomical Data
Center (ADC).

\acknowledgments

We thank Guy Worthey for providing the theoretical isochrones under BATC
colors, David Marcus for his help with the BATC filter design, Sam Pascarelle
for the calibrated MMT night sky spectrum, and Jeremy Goodman for helpful
discussion. We also thank the referee for valuable comments and suggestions.
This research is supported in part by an International U.S. NSF Grant
INT-93-01805, by the National Science Foundation of China, by Arizona State
University, the University of Arizona and Western Connecticut State University
and by the National Science Council of Taiwan under the grant
NSC84-2112-M-008-024.

\newpage
\oddsidemargin 0mm
\begin{center}
\begin{large}
{\bf Figure Captions}
\end{large}
\end{center}
Figure 1. Transmission curves of the BATC filters superimposed on a typical
sky spectrum.  The Number 10 filter (7215$\rm \AA$) was used at the time when 
the M67 data were taken.  Later, this passband was divided among filters 10a
(7050$\rm \AA$) and 10b filter (7490$\rm \AA$). Note that most of the 
important sky lines are avoided by the BATC filter bandpasses.

\vspace*{2mm}

Figure 2. The central $58' \times 58'$ field of view of the BATC CCD, centered
on M67.  North is up, East is left.  The scale of the image is given by the 
arrow at the lower-right corner. Numbers at the bottom, left, top and right 
sides, respectively, are R.A. and Dec. (2000.0) and the angles of azimuth
and altitude of the field at the time of observation. 

\vspace*{2mm}

Figure 3. Photometric precision in 9 filters computed from five shift-field
exposusre. Delta mag is the magnitude difference between a single observation 
and the average of five observations.  The thick solid line is the average of 
observed scatter while the thick dashed line is the average of predicitions 
given by DAOPHOT II.  Note the DAOPHOT II predictions somewhat overestimate
the $1 \sigma$ errors actually observed.

\vspace*{2mm}

Figure 4. The relationship between BATC magnitudes and UBVRI magnitudes. 
BATC magnitudes are compared with BVI observation of MMJ (1993) (left hand
side of figure) and the BVR observations of Gilliland et al. (1991) (right
hand side of figure).  The scatter in these correlations is typically 0.02 mag. 

\vspace*{2mm}

Figure 5. The internal astrometric precision of the BATC data, computed from 
five shift-field exposures. The solid line is the 1$\sigma$ scatter of 
the observations. 

\vspace*{2mm}

Figure 6. The $\rm (m_{3890} - m_{6075})$ vs. $\rm m_{3890}$ color magnitude
diagram (CMD) for M67.  The upper graph is for the whole field observed around
M67; the lower graph is only for the stars within 15$'$ of the cluster center.
Stars with different membership probabilities (Girard et al. 1989) are marked
with different symbols: $>$ 90\%: plus signs;  80\% $\leq $90\%, open circles;
50\% $\leq$ 80\%: stars; $<$ 50\%: crosses.  Stars with no measured membership
probabilities are plotted as points.  This CMD gives us the cleanest view of
the main sequence of M67.

\vspace*{2mm}

Figure 7. The $\rm (m_{3890} - m_{9190})$ vs. $\rm m_{3890}$ CMD. Symbols are
the same as in Figure~6.  This CMD gives us the widest passband colors that
also go to the faintest magnitudes.

\vspace*{2mm}

Figure 8. The $\rm (m_{6075} - m_{7215})$ vs. $\rm m_{6075}$ CMD. Symbols are
the same as in Figure~6.  This CMD goes the faintest of all the BATC filter
observations.

\vspace*{2mm}

Figure 9. The $\rm (m_{3890} - m_{6075})$ vs. $\rm m_{3890}$ CMD, now showing
only those stars with membership probability $>$ 80\%.

\vspace*{2mm}

Figure 10. Worthey-VandenBerg-Kurucz isochrone models fit to the observed $\rm
(m_{3890} - m_{6075})$ vs. $\rm m_{3890}$ CMD.  $\rm (m-M)_{0} = 9.47$ and
E(B--V) = 0.05 are assumed; see text for details. The values of age and [Fe/H]
of each isochrone are shown in the graphs.  Data for stars with known
membership probabilities $\ge$80\% are plotted as open circles; all other
stars are plotted as dots.

\vspace*{2mm}

Figure 11. Worthey-VandenBerg-Kurucz isochrone models fit to the observed $\rm
(m_{6075} - m_{7215})$ vs. $\rm m_{6075}$ CMD.  Same models, distance modulus
and [Fe/H] as for Figure~10.

\vspace*{2mm}

Figure 12. (a) Tracks of iso-primary-mass binaries and iso-mass-ratio binaries
in the $\rm (m_{3890} - m_{6075})$ vs. $\rm m_{3890}$ CMD. Iso-primary-mass
tracks are plotted as solid lines in steps 0.02 in the primary mass;
iso-mass-ratio tracks as dotted lines in steps of 0.05 in seconday/primary
mass ratio. (b) Same tracks plotted over the observed M67 $\rm (m_{3890} - 
m_{6075})$ vs. $\rm m_{3890}$ CMD; data for stars with membership probabilities
$>$80\% are plotted as open circles.

\vspace*{2mm}

Figure 13. The results of our ``toy'' model of a stellar population of age
and metallicity similar to M67 that has 50\% of its stars being binaries 
with a random distribution of binary mass-ratios. (a) The predicted
distribution of stars in the $\rm (m_{3890} - m_{6075})$ vs. $\rm m_{3890}$ CMD.
(b) The predicted distribution of stars in the $\rm (m_{6075} - m_{7215})$ vs. 
$\rm m_{6075}$ CMD.

\vspace*{2mm}

Figure 14. The distribution of stars relative to single star MS for real data
(solid) compared to that obtained from the ``toy'' model (dashed).  Only stars
with colors $\rm (m_{3890}-m_{6075})$ between 1.4 and 2.5 are used for this
histogram, both model and real data. The fiducial single star MS is defined as
the 4 Gyr, [Fe/H] = --0.10 Worthey-VandenBerg-Kurucz model from Figure~10. 
The apparent bimodal distribution of stars, with one peak at the single star
isochrone and the other 0.7 mag brighter, is evident.  However, the asymmetric
shape of the ``single'' star peak, which scatters preferentially towards
brighter mags and redder colors, implies the existence of large numbers of
low-mass-ratio binary systems in M67 masquerading as apparent single stars. 
Note that the real star distribution includes all stars in this magnitude
interval, known member or not.

\vspace*{2mm}

Figure 15. The apparent magnitude distribution of field stars, as obtained
within an annulus of 37.5$'$ to 42.5$'$ in radius from the M67 cluster center. 
The dashed line is the predicted magnitude distribution of the Bahcall \& 
Soneira (1980) model. 

\vspace*{2mm}

Figure 16. The observed cumulative radial distributions for seven kinds
of M67 stars:  subgiants+giants (light solid line); three absolute magnitude 
ranges of MS stars --- $13.8<\rm m_{3890}\leq14.5$ (heavy solid line),
$14.5<\rm m_{3890}\leq15.6$ (long dashed line) and $15.6<\rm m_{3890}\leq18.5$ 
(dark dot-dashed line); blue stragglers (light dotted line), and two absolute 
magnitude ranges of photometric binary stars --- $14.5<\rm m_{3890}\leq15.6$ 
(light dot-dashed line) and $15.6<\rm m_{3890}\leq18.5$ (dash-dot-dot-dotted line).

\vspace*{2mm}

Figure 17. The $\rm (m_{3890} - m_{6075})$ vs. $\rm m_{3890}$ CMD as observed
in nine different positions centered around M67.  The M67 MS stands out
clearly along the upper-left (NE) to lower-right (SW) diagonal, yielding an
angle of $15^{\circ}$ relative to galactic plane) and an an elliptical shape
for this cluster.

\vspace*{2mm}

Figure 18. The observed 7215$\rm \AA$ luminosity function of M67 separated
into three volumes of increasing size: (a) to a radius of 5$'$; (b) to a
radius of 16.67$'$ and (c) to a radius of 33.33$'$.  See text for a full
discussion of this figure.

\vspace*{2mm}

Figure 19. The observed mass function for the whole cluster, as derived from
the luminosity function of Figure 18c, combined with the isochrone fit of the
Worhey-VandenBerg-Kurucz model.  The lower mass limit is 0.5 $\rm M_\odot$
and the dashed line is a $\eta = -2.0$ Salpeter law fit to the stars with
masses 1.2 to 0.8 $\rm M_\odot$.

\newpage
Table 1. The BATC Intermediate-Band Filter Set\\

\begin{center}
\vspace*{4mm}
\begin{tabular}{cccc} \hline \hline
 Filter No. & CW(\AA) & FWHM(\AA) &  Observed ?  \\ \hline
   1    & 3360  & 360   & No   \\
   2    & 3890  & 340   & Yes  \\
   3    & 4210  & 320   & No   \\
   4    & 4550  & 340   & No   \\
   5    & 4925  & 390   & No   \\
   6    & 5270  & 340   & No   \\
   7    & 5795  & 310   & Yes  \\
   8    & 6075  & 310   & Yes  \\
   9    & 6660  & 480   & Yes  \\
 10a    & 7050  & 300   & No   \\
  10    & 7215  & 550   & Yes  \\
 10b    & 7490  & 330   & No   \\
  11    & 8020  & 260   & Yes  \\
  12    & 8480  & 180   & Yes  \\
  13    & 9190  & 260   & Yes  \\
  14    & 9745  & 270   & Yes  \\ \hline \hline
\end{tabular}
\end{center}
\newpage
Table 2 : Observation Log, 7Jan94 -- 2Feb94
\vspace*{4mm}
\begin{center}
\begin{tabular}{ccccccccccc} \hline \hline
Date & 3890\AA & 5795\AA & 6075\AA & 6660\AA & 7215\AA & 8020\AA & 8480\AA
& 9190\AA & 9745\AA & FWHM  \\
(yymmdd) &(sec)&(sec)&(sec)&(sec)&(sec)&(sec)&(sec)&(sec)&(sec)&(arcsec) \\
\tableline
 940107      &  &  180 &  &  180 &&  180 &&      &  360 &   2.1 \\
 940109      &  & 1200 &  & 1200 && 1200 &&      & 3600 &   3.2 \\
 940117 (I)  &  &  300 &  &  300 &&  300 &&      & 1200 &   6.0 \\
 940117 (II) &  &  180 &  &  180 &&  180 &&      &  600 &   4.5 \\
 940117 (III)&  &  180 &  &  180 &&  180 &&      &  600 &   5.6 \\
 940117 (IV) &  &  180 &  &      &&  180 &&      &      &   6.1 \\ \hline
 940202      &  900 &  &  300 &   &  300 &&  600 &  600 & & 3.0 \\
 940204 (I)  &  600 &  &  180 &   &  180 &&  360 &  360 & & 2.7 \\
 940204 (II) &  600 &  &  180 &   &  180 &&  360 &  360 & & 2.1 \\
 940204 (III)&  600 &  &  180 &   &  180 &&  360 &  360 & & 1.7 \\
 940204 (IV) &  600 &  &  180 &   &  180 &&  360 &  360 & & 1.7 \\
 940205      & 2400 &  & 1200 &   & 1200 && 2400 & 2400 & & 2.6 \\ \hline
 Total       & 5700 & 2220 & 2220 & 2040 & 2220 & 2220 & 4440 & 4440 & 6360 &\\
$\rm m_{\rm lim}$ & 19.6 & 19.5 & 20.0 & 20.0 & 20.1 & 18.4 & 18.0 & 18.6
& 16.6 &\\ \hline \hline
\end{tabular}
\end{center}
\newpage
Table 3 : Photometric Data

\begin{center}
\vspace*{4mm}
\begin{tabular}{cccccccc}\hline \hline
Filter & Num Stars & Num Obs & a & $\rm \sigma(a)$ & b & $\rm \sigma(b)$
& residual\\
$\rm \AA$ & & & & & & & \\\hline
3890  & 3 &  7 & 2.663 & 0.021 & 0.528 & 0.015 & 0.015 \\
6075  & 3 &  7 & 1.662 & 0.017 & 0.207 & 0.012 & 0.007 \\
7215  & 3 &  7 & 1.135 & 0.028 & 0.139 & 0.019 & 0.010 \\
8480  & 3 &  7 & 3.404 & 0.056 & 0.065 & 0.038 & 0.015 \\
9190  & 3 &  7 & 3.982 & 0.047 & 0.089 & 0.032 & 0.014 \\ \hline
5795  & 4 &  4 & 1.644 & 0.031 & 0.285 & 0.020 & 0.013 \\
6660  & 3 &  4 & 0.967 & 0.063 & 0.243 & 0.048 & 0.021 \\
8020  & 4 &  4 & 2.300 & 0.025 & 0.116 & 0.014 & 0.006 \\
9745  & 4 &  4 & 4.680 & 0.037 & 0.146 & 0.023 & 0.012 \\\hline \hline
\end{tabular}
\end{center}
\newpage
\oddsidemargin -7mm
Table 4 : BATC Secondary Standard Stars in M67
\begin{center}
\vspace*{4mm}
\begin{scriptsize}  
\begin{tabular}{cccccccccccccc} \hline \hline
ID &MMJ\# & Sanders\#& RA(2000) & DEC(2000)&3890 &5795 & 6075 & 6660 & 7215 
& 8020 & 8480 & 9190 & 9745  \\ 
\tableline
 &&&&&&&&&&&&&  \\
4118 & 6505 & 1288 & 8 51 42.36 & 11 51 22.87 & 13.449 & 11.103 & 10.887 & 10.676 & 10.609 & 10.596 & 10.465 & 10.341 & 10.316 \\
3903 & 6504 & 1281 & 8 51 34.26 & 11 50 54.31 & 14.578 & 13.621 & 13.500 & 13.374 & 13.392 & 13.422 & 13.365 & 13.311 & 13.288 \\
3751 & 5871 & 2223 & 8 51 29.86 & 11 51 29.76 & 14.060 & 13.250 & 13.132 & 13.025 & 13.055 & 13.097 & 13.053 & 12.985 & 12.973 \\
3726 & 6503 & 1279 & 8 51 28.99 & 11 50 33.01 & 12.822 & 10.366 & 10.137 & \nodata& \nodata&  9.847 &  9.714 &  9.579 &  9.558 \\
3347 & 6489 & 1054 & 8 51 17.05 & 11 50 46.26 & 13.342 & 10.977 & 10.765 & 10.549 & 10.424 & 10.459 & 10.338 & 10.213 & 10.173 \\
3254 & 5559 & 2221 & 8 51 14.44 & 11 50 40.13 & 14.313 & 13.308 & 13.183 & 13.050 & 13.064 & 13.110 & 13.047 & 12.976 & 12.964 \\
3202 & 5534 & 1052 & 8 51 12.68 & 11 50 34.43 & 14.567 & 13.558 & 13.424 & 13.307 & 13.315 & 13.334 & 13.294 & 13.227 & 13.208 \\
3283 & 5573 & 1049 & 8 51 15.34 & 11 50 14.17 & 13.703 & 12.735 & 12.603 & 12.478 & 12.490 & 12.541 & 12.468 & 12.406 & 12.408 \\
3659 & 3659 &  103 & 8 51 26.52 & 11 49 20.14 & 14.420 & 13.405 & 13.277 & 13.161 & 13.177 & 13.213 & 13.136 & 13.083 & 13.068 \\
3607 & 5781 & 1027 & 8 51 24.95 & 11 49  0.65 & 14.226 & 13.169 & 13.036 & 12.914 & 12.927 & 12.958 & 12.882 & 12.827 & 12.811 \\
3541 & 5739 & 1024 & 8 51 22.91 & 11 48 49.28 & 13.600 & 12.625 & 12.502 & 12.397 & 12.392 & 12.447 & 12.364 & 12.306 & 12.310 \\
3668 & 6487 &  102 & 8 51 26.84 & 11 48 40.37 & 11.444 & 10.415 & 10.242 & 10.163 &\nodata & 10.217 & 10.134 & 10.079 & 10.084 \\
3717 & 5844 & 1011 & 8 51 28.70 & 11 48  2.04 & 14.868 & 13.718 & 13.575 & 13.437 & 13.434 & 13.437 & 13.357 & 13.281 & 13.259 \\
3536 & 6485 & 1010 & 8 51 22.80 & 11 48  1.66 & 12.724 & 10.293 & 10.066 &  9.928 &\nodata &  9.808 &  9.661 &  9.534 &  9.517 \\
3349 & 6486 & 1016 & 8 51 17.10 & 11 48 16.01 & 12.971 & 10.087 &  9.750 &\nodata &\nodata &  9.463 &  9.317 &  9.177 &  9.124 \\
3186 & 5522 & 1021 & 8 51 12.23 & 11 48 34.62 & 14.885 & 13.828 & 13.697 & 13.580 & 13.584 & 13.627 & 13.552 & 13.489 & 13.458 \\
3755 & 6499 & 1250 & 8 51 29.91 & 11 47 16.78 & 12.565 &  9.433 &\nodata &  8.912 &\nodata &\nodata &  8.547 &  8.403 &  8.375 \\
3288 & 5571 & 1005 & 8 51 15.45 & 11 47 31.34 & 13.470 & 12.616 & 12.497 & 12.405 & 12.417 & 12.483 & 12.407 & 12.369 & 12.364 \\
3267 & 5562 & 1003 & 8 51 14.76 & 11 47 23.87 & 13.738 & 12.759 & 12.624 & 12.520 & 12.521 & 12.570 & 12.495 & 12.437 & 12.433 \\
3449 & 5667 &  997 & 8 51 19.91 & 11 47  0.32 & 12.834 & 12.071 & 11.942 & 11.869 & 11.880 & 11.957 & 11.890 & 11.853 & 11.834 \\
3458 & 5675 & 2205 & 8 51 20.33 & 11 45 52.37 & 14.084 & 13.092 & 12.955 & 12.849 & 12.849 & 12.888 & 12.814 & 12.764 & 12.746 \\
3465 & 5688 & 2204 & 8 51 20.57 & 11 46 16.31 & 13.608 & 12.843 & 12.732 & 12.653 & 12.679 & 12.745 & 12.683 & 12.653 & 12.633 \\
3462 & 6483 &  988 & 8 51 20.55 & 11 46  4.74 & 14.155 & 13.114 & 12.973 & 12.857 & 12.865 & 12.909 & 12.834 & 12.777 & 12.778 \\
3458 & 5678 & 2205 & 8 51 20.33 & 11 45 52.37 & 14.084 & 13.092 & 12.955 & 12.849 & 12.849 & 12.888 & 12.814 & 12.764 & 12.746 \\
3380 & 5624 &  986 & 8 51 17.99 & 11 45 54.11 & 13.637 & 12.658 & 12.524 & 12.406 & 12.414 & 12.463 & 12.384 & 12.332 & 12.318 \\
3805 & 5896 & 1234 & 8 51 31.24 & 11 45 50.54 & 13.538 & 12.549 & 12.415 & 12.306 & 12.300 & 12.357 & 12.267 & 12.214 & 12.218 \\
3472 & 5695 &  976 & 8 51 20.81 & 11 45  2.39 & 14.096 & 13.017 & 12.883 & 12.766 & 12.760 & 12.806 & 12.730 & 12.681 & 12.661 \\
3176 & 3176 &  977 & 8 51 11.78 & 11 45 22.05 &  9.835 & 10.096 & 10.048 & 10.148 & \nodata& 10.500 & 10.471 & 10.480 & 10.526 \\
3083 & 5464 &  990 & 8 51  8.64 & 11 46 11.71 & 14.325 & 13.362 & 13.227 & 13.122 & 13.133 & 13.171 & 13.087 & 13.052 & 13.025 \\ \hline \hline
\end{tabular}
\end{scriptsize}  
\end{center}
\newpage
Table 5 : Best Fit Isochrone of M67 in BATC Filter System

\begin{center}
\vspace*{4mm}
\begin{tabular}{cccc} \hline \hline
Mass ($\rm M_{\odot}$) & $\rm M_{3890}$ & $\rm M_{6075}$ & $\rm M_{7215}$ \\  \hline
0.52  &  11.95  &  8.71  & 8.31\\
0.57  &  11.39  &  8.22  & 7.84\\
0.63  &  10.83  &  7.77  & 7.42\\
0.67  &  10.24  &  7.35  & 7.03\\
0.71  &  9.64   &  6.96  & 6.68\\
0.75  &  9.04   &  6.59  & 6.35\\
0.79  &  8.46   &  6.25  & 6.04\\
0.83  &  7.90   &  5.91  & 5.73\\
0.87  &  7.34   &  5.57  & 5.42\\
0.92  &  6.77   &  5.21  & 5.09\\
0.98  &  6.21   &  4.84  & 4.74\\
1.04  &  5.64   &  4.44  & 4.35\\
1.11  &  5.05   &  3.98  & 3.92\\
1.20  &  4.49   &  3.47  & 3.41\\
1.23  &  4.08   &  3.12  & 3.08\\
1.26  &  3.88   &  2.84  & 2.78\\
1.28  &  3.89   &  2.73  & 2.65\\
1.29  &  4.23   &  2.77  & 2.65\\
1.30  &  4.99   &  3.03  & 2.85\\
1.31  &  4.81   &  2.53  & 2.32\\ \hline \hline 
\end{tabular}
\end{center}
\newpage
Table 6 : Spatial Distribution of Stars within M67
\begin{center}
\vspace*{4mm}
\begin{tabular}{ccccc} \hline \hline
type of star  &$\rm m_{3890}$ & $<Mass>$  & Half-mass radius
& half-mass radius \\
 && $\rm M_{\odot}$ & arcmin & pc \\ \hline
blue straggler   &              & $\geq 2.00$ &  $7.16 \pm 2.13$ & $1.63 \pm 0.49$ \\
binary sequence  & 14.5 -- 15.6 & $\leq 1.96$ &  $7.68 \pm 2.07$ & $1.75 \pm 0.44$ \\
binary sequence  & 15.6 -- 18.5 & $\leq 1.62$ &  $9.20 \pm 1.69$ & $2.10 \pm 0.39$ \\ \hline 
giant + subgiant &              & $\geq 1.30$ &  $7.70 \pm 0.84$ & $1.75 \pm 0.19$ \\ \hline
MS    & 13.8 -- 14.5 & $\geq 1.20$ & $10.23 \pm 1.65$ & $2.33 \pm 0.38$ \\ 
MS    & 14.5 -- 15.6 & $\geq 1.07$ & $11.03 \pm 1.18$ & $2.51 \pm 0.27$ \\
MS    & 15.6 -- 18.5 & $\geq 0.87$ & $11.81 \pm 1.00$ & $2.69 \pm 0.23$ \\ \hline 
MS b  & 15.6 -- 18.5 & $\geq 0.87$ & $12.00 \pm 1.35$ & $2.73 \pm 0.31$ \\
MS r  & 15.6 -- 18.5 & $\geq 0.87$ & $11.29 \pm 2.02$ & $2.57 \pm 0.46$ \\ \hline\hline
\end{tabular}
\end{center}
\newpage
Table 7 : Radial-Dependent Mass Function of M67
\begin{center}
\vspace*{4mm}
\begin{tabular}{cccc} \hline \hline
radius & $\eta$ & total mass (no binary correction)
& total mass (binary corrected) \\
arcsec & & $\rm (M > 0.50 M_{\odot})$ & $\rm (M > 0.50 M_{\odot})$ \\ \hline
300  & $+1.12\pm0.04$ &  218 &  272 \\
500  & $+0.76\pm0.28$ &  361 &  451 \\
1000 & $+0.01\pm0.07$ &  672 &  840 \\
1500 & $-0.34\pm0.08$ &  931 & 1164 \\
2000 & $-0.58\pm0.14$ & 1016 & 1270 \\ \hline \hline
\end{tabular}
\end{center}
\begin{figure}
\newpage
\vspace{-6cm}

\epsfysize=600pt \epsfbox{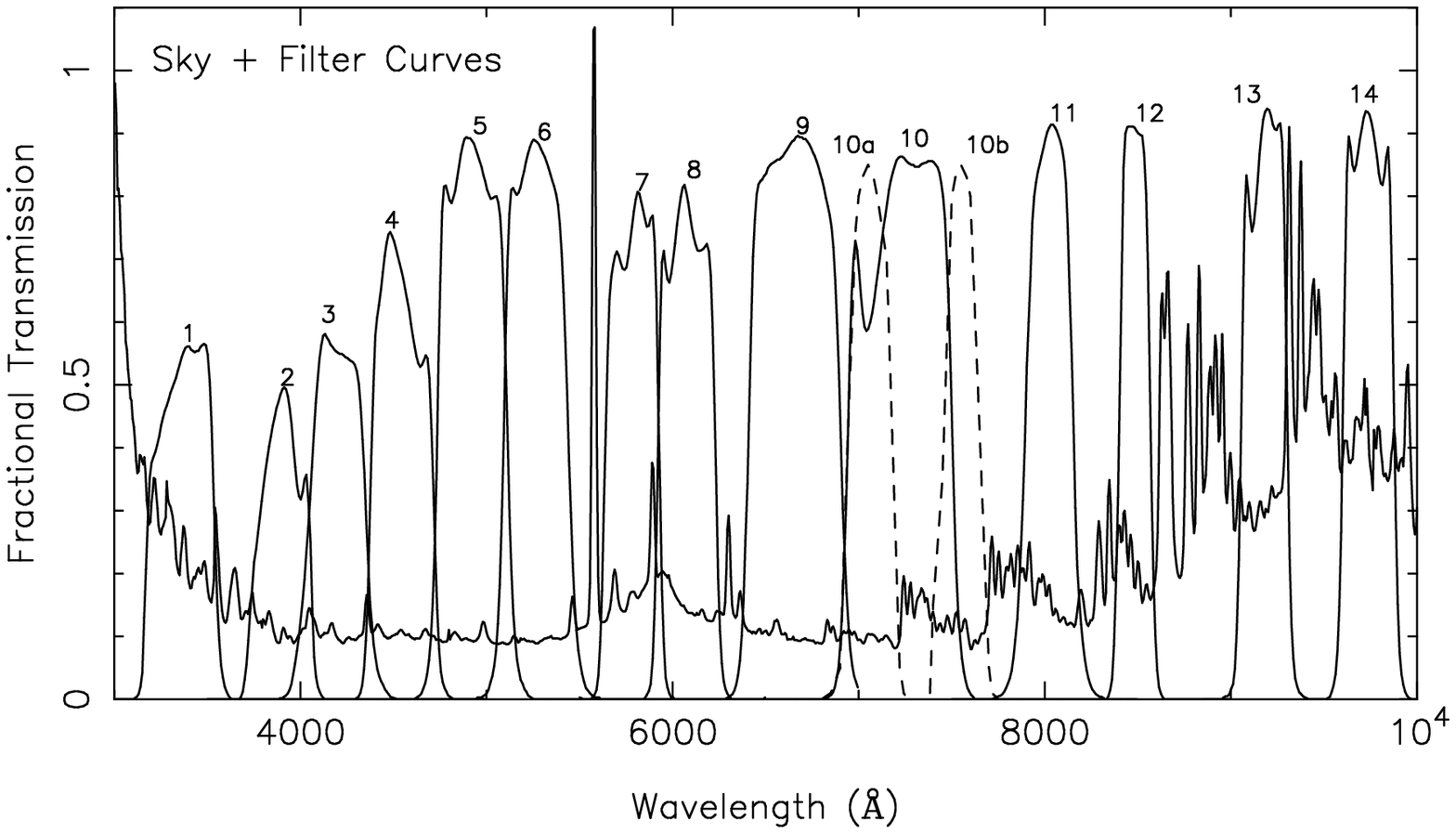}

\vspace{6.5cm}
\hspace{12cm}Fig 1. Fan et al.
\end{figure}
\begin{figure}
\epsfysize=600pt \epsfbox{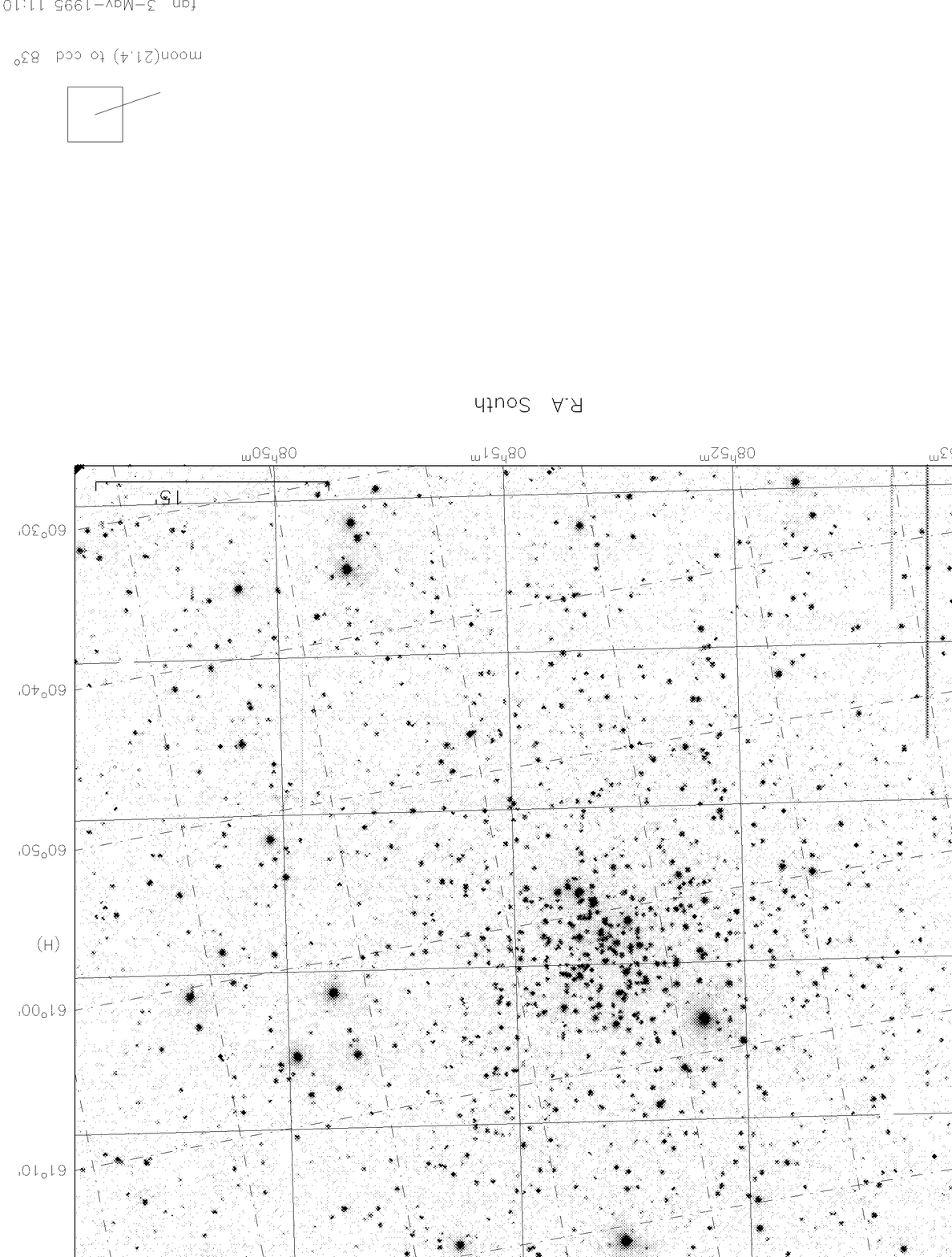}

\vspace{1cm}
\hspace{12cm}Fig 2. Fan et al.
\end{figure}
\begin{figure}
\epsfysize=600pt \epsfbox{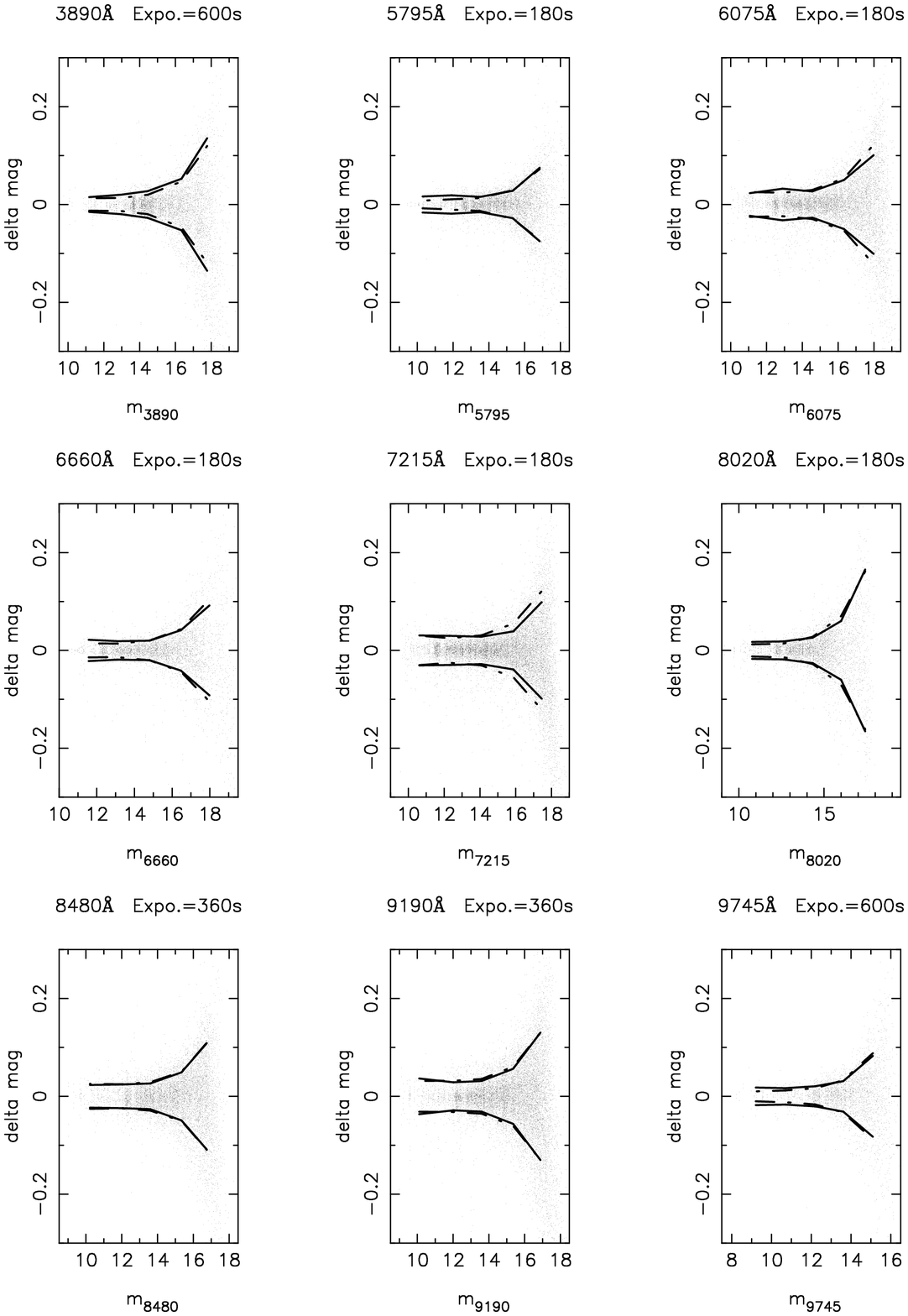}

\vspace{1cm}
\hspace{12cm}Fig 3. Fan et al.
\end{figure}
\begin{figure}
\epsfysize=600pt \epsfbox{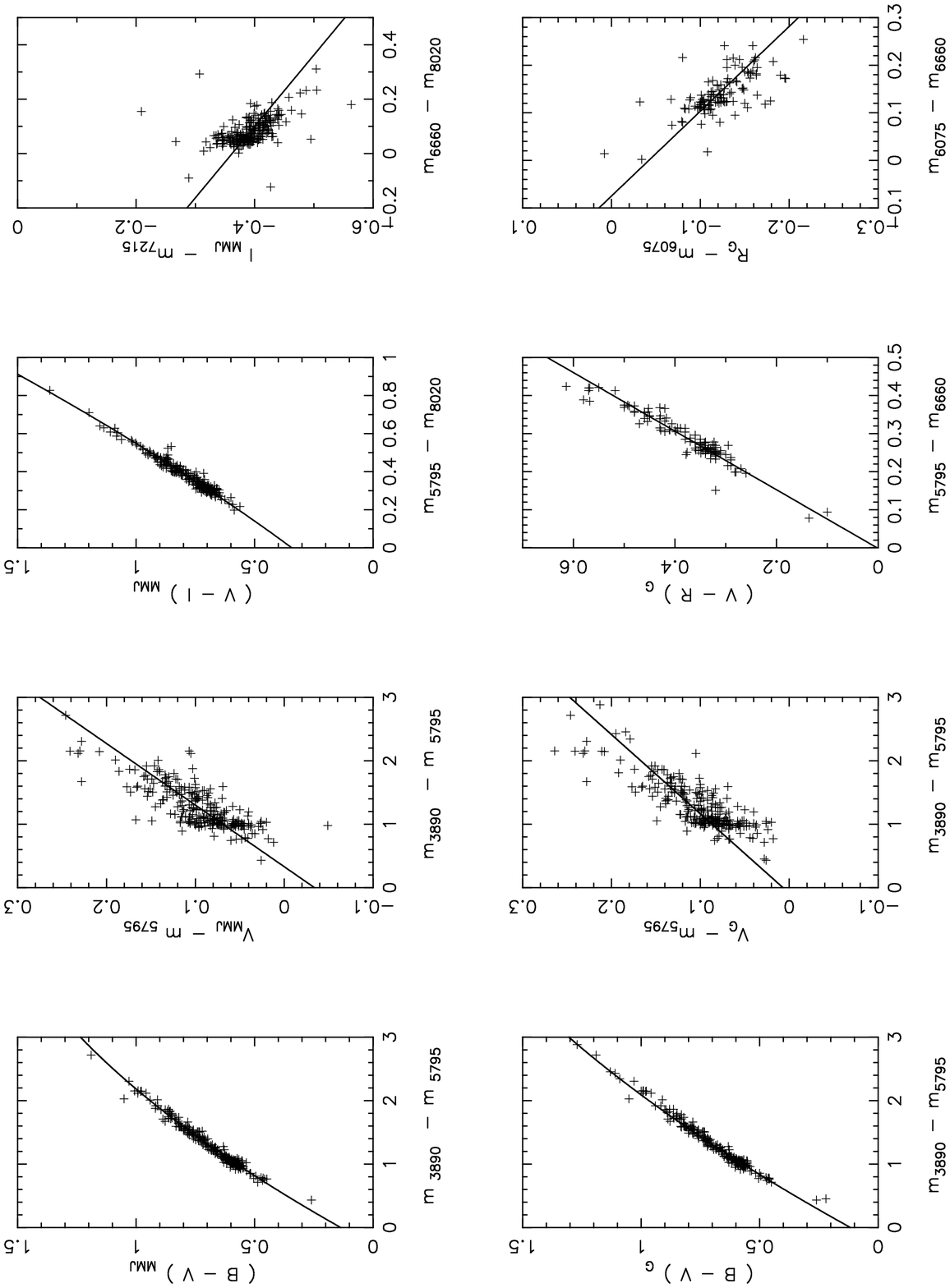}

\vspace{1cm}
\hspace{12cm}Fig 4. Fan et al.
\end{figure}
\begin{figure}
\vspace{-6cm}

\epsfysize=600pt \epsfbox{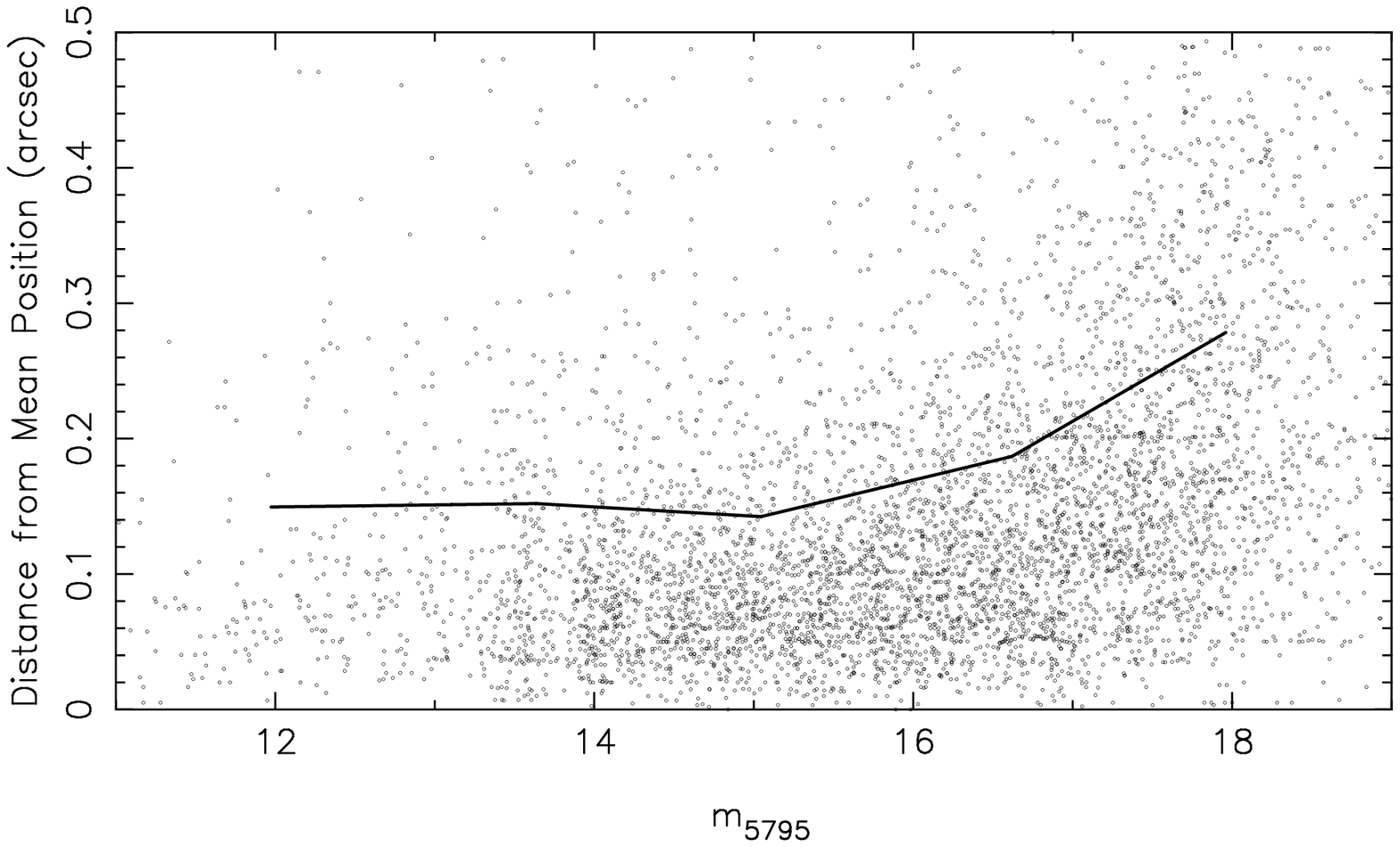}

\vspace{6.5cm}
\hspace{12cm}Fig 5. Fan et al.
\end{figure}
\begin{figure}
\epsfysize=600pt \epsfbox{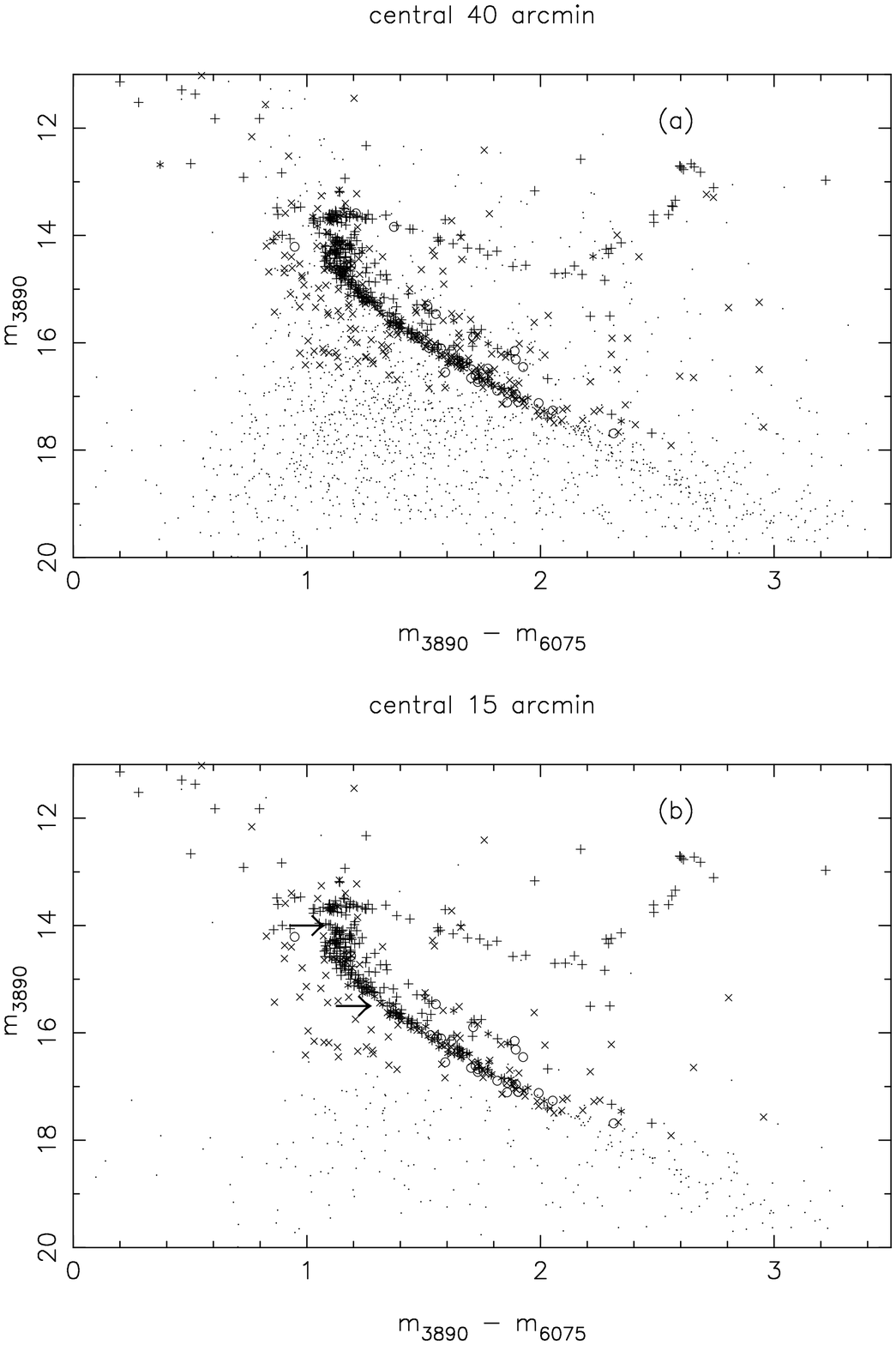}

\vspace{1cm}
\hspace{12cm}Fig 6. Fan et al.
\end{figure}
\begin{figure}
\epsfysize=600pt \epsfbox{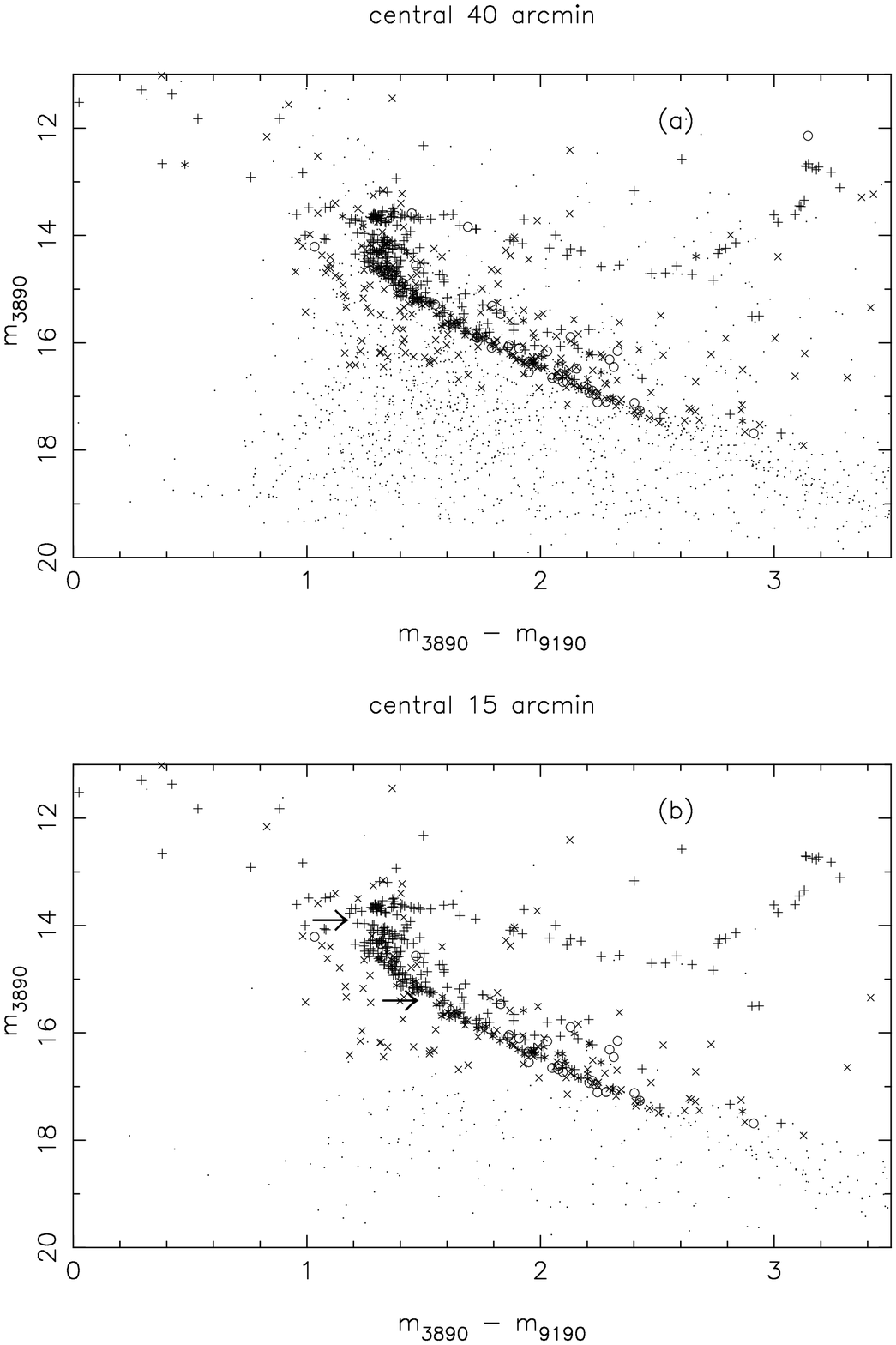}

\vspace{1cm}
\hspace{12cm}Fig 7. Fan et al.
\end{figure}
\begin{figure}
\epsfysize=600pt \epsfbox{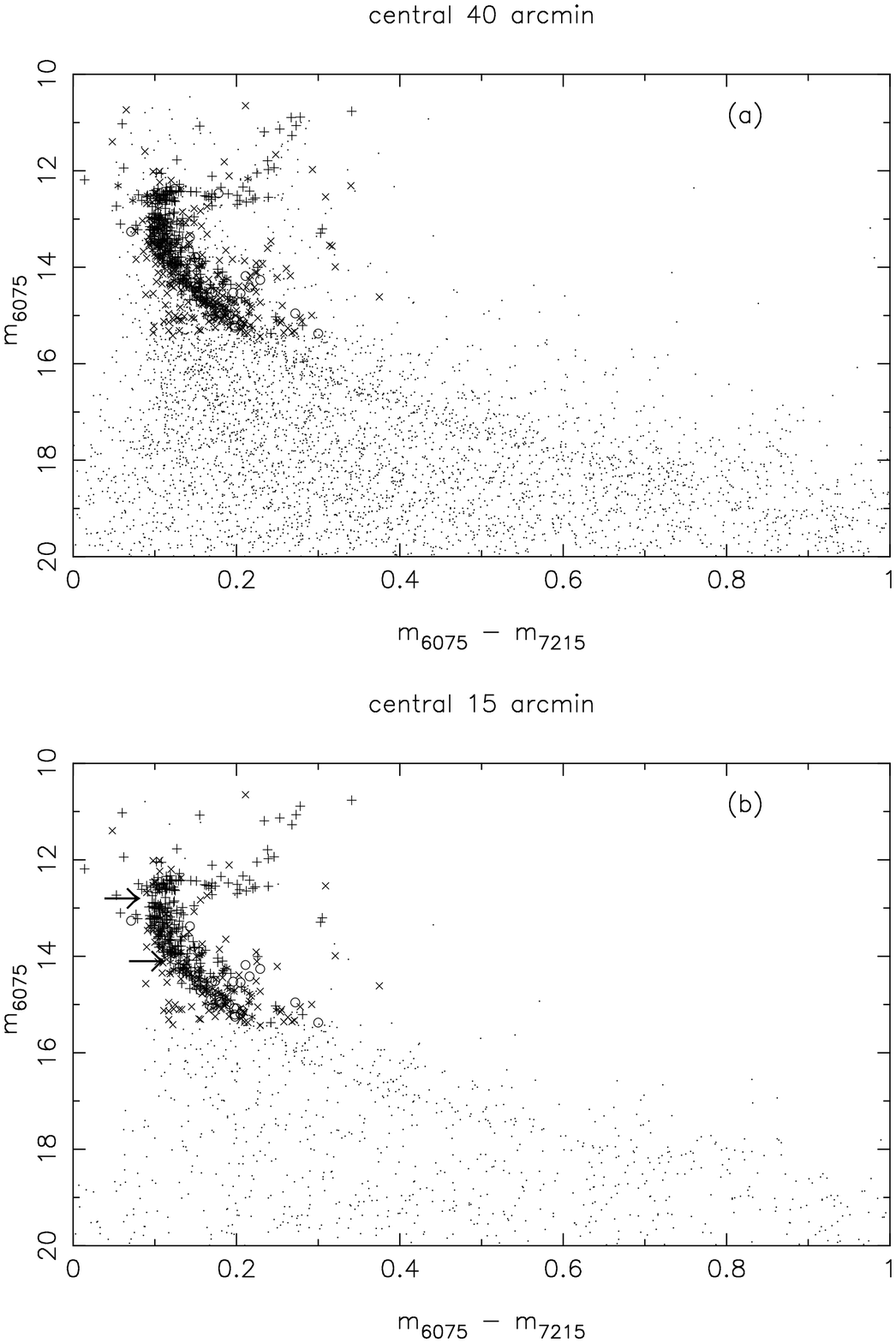}

\vspace{1cm}
\hspace{12cm}Fig 8. Fan et al.
\end{figure}
\begin{figure}
\epsfysize=600pt \epsfbox{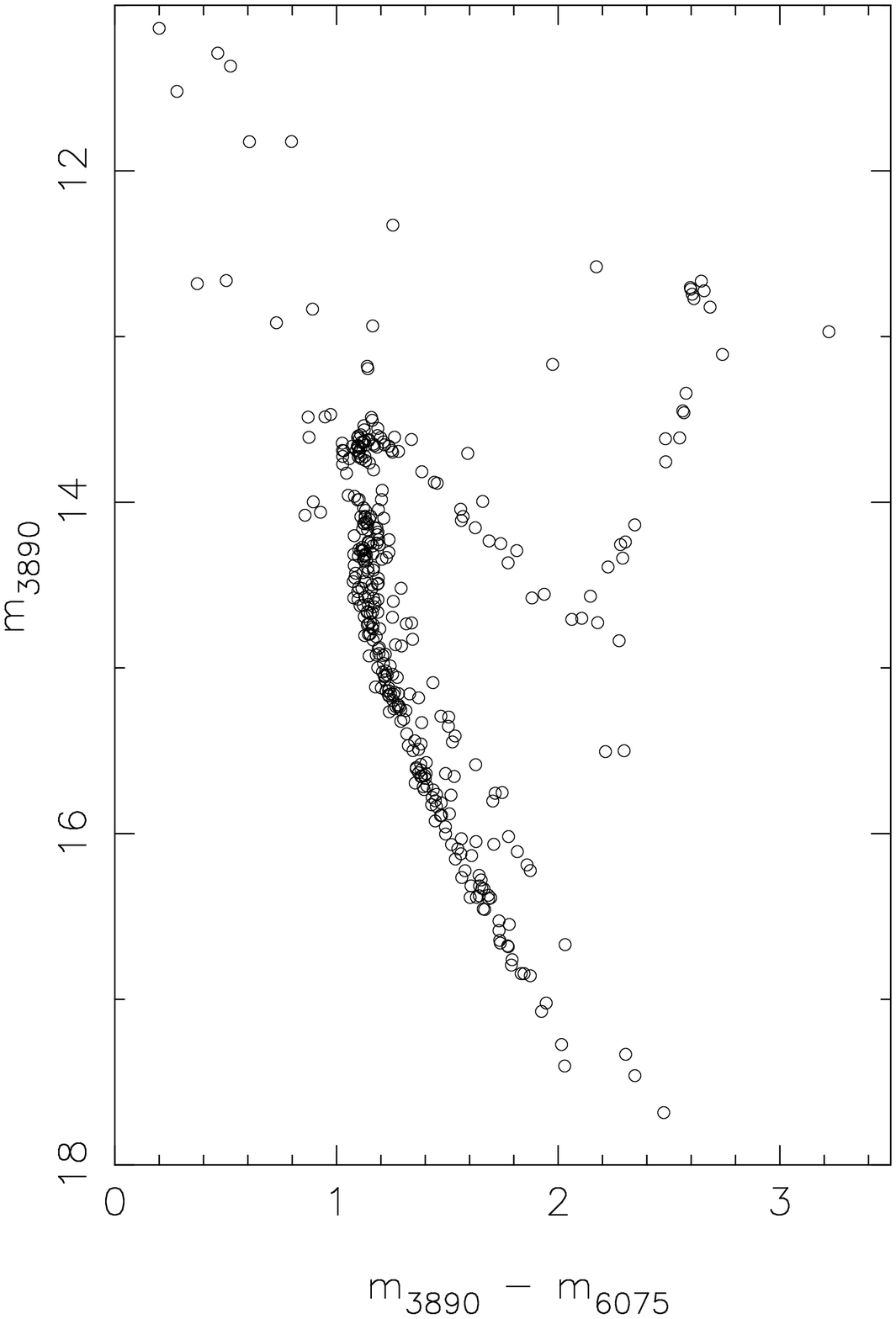}

\vspace{1cm}
\hspace{12cm}Fig 9. Fan et al.
\end{figure}
\begin{figure}
\epsfysize=600pt \epsfbox{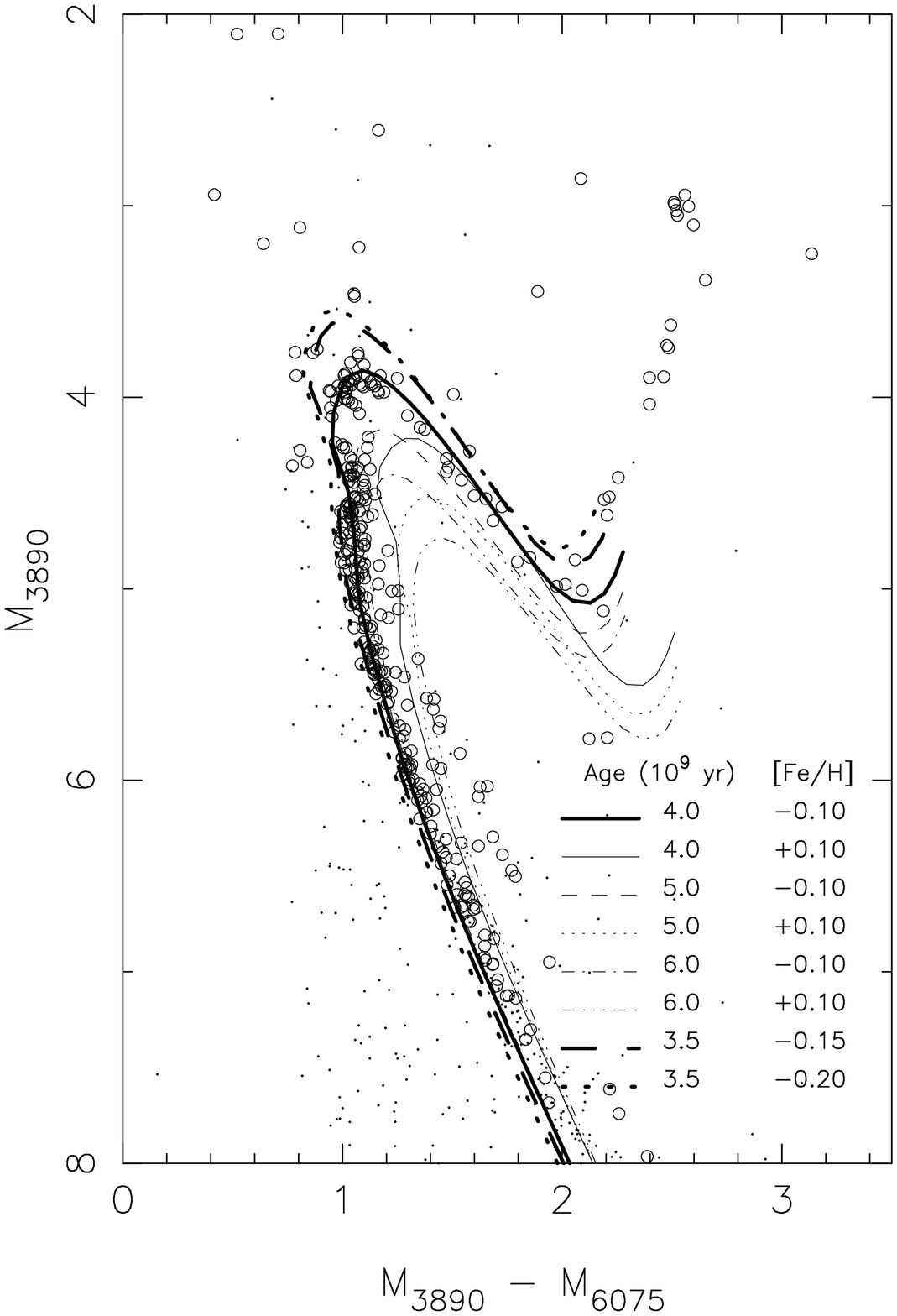}

\vspace{1cm}
\hspace{12cm}Fig 10. Fan et al.
\end{figure}
\begin{figure}
\epsfysize=600pt \epsfbox{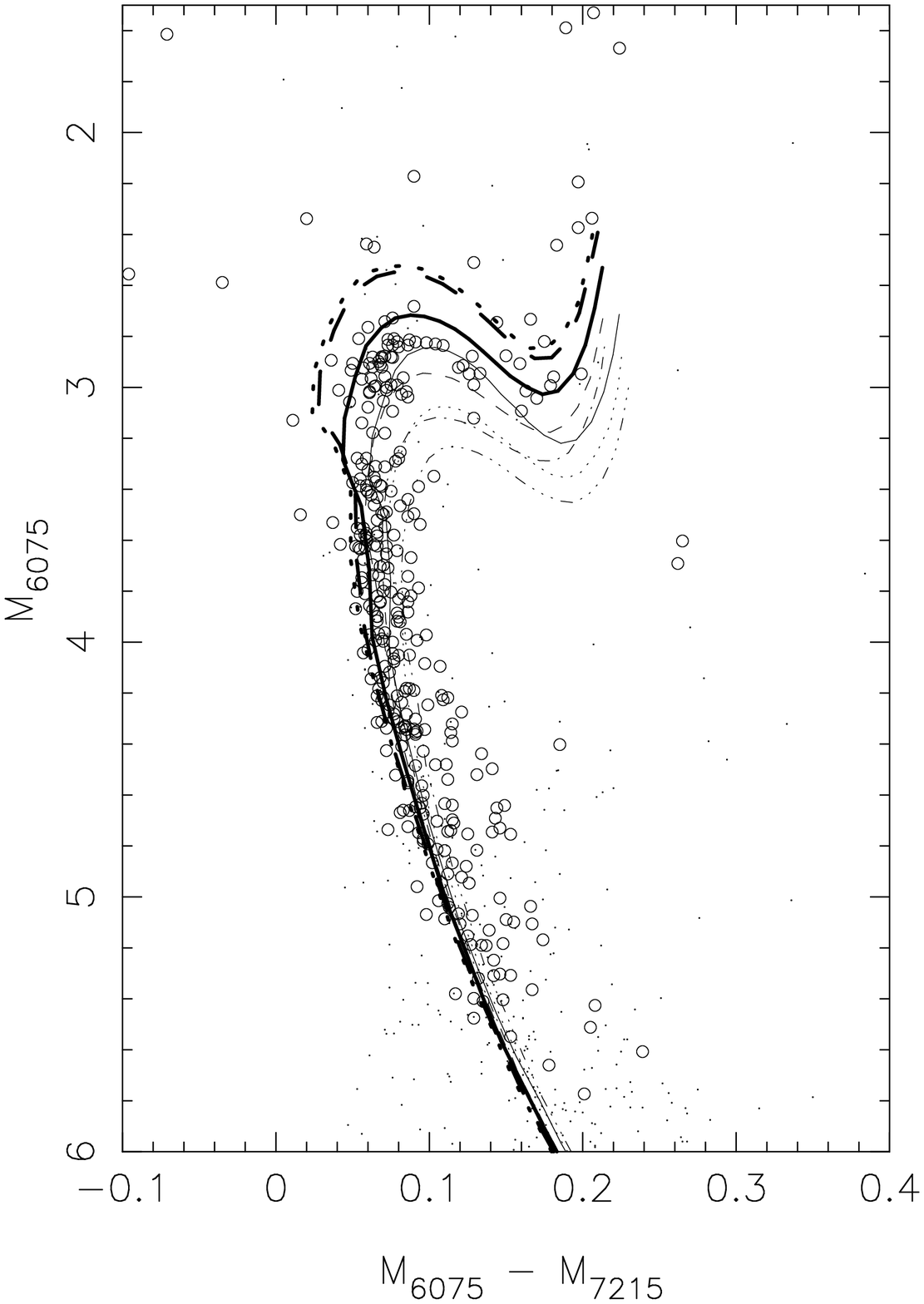}

\vspace{1cm}
\hspace{12cm}Fig 11. Fan et al.
\end{figure}
\begin{figure}
\epsfysize=600pt \epsfbox{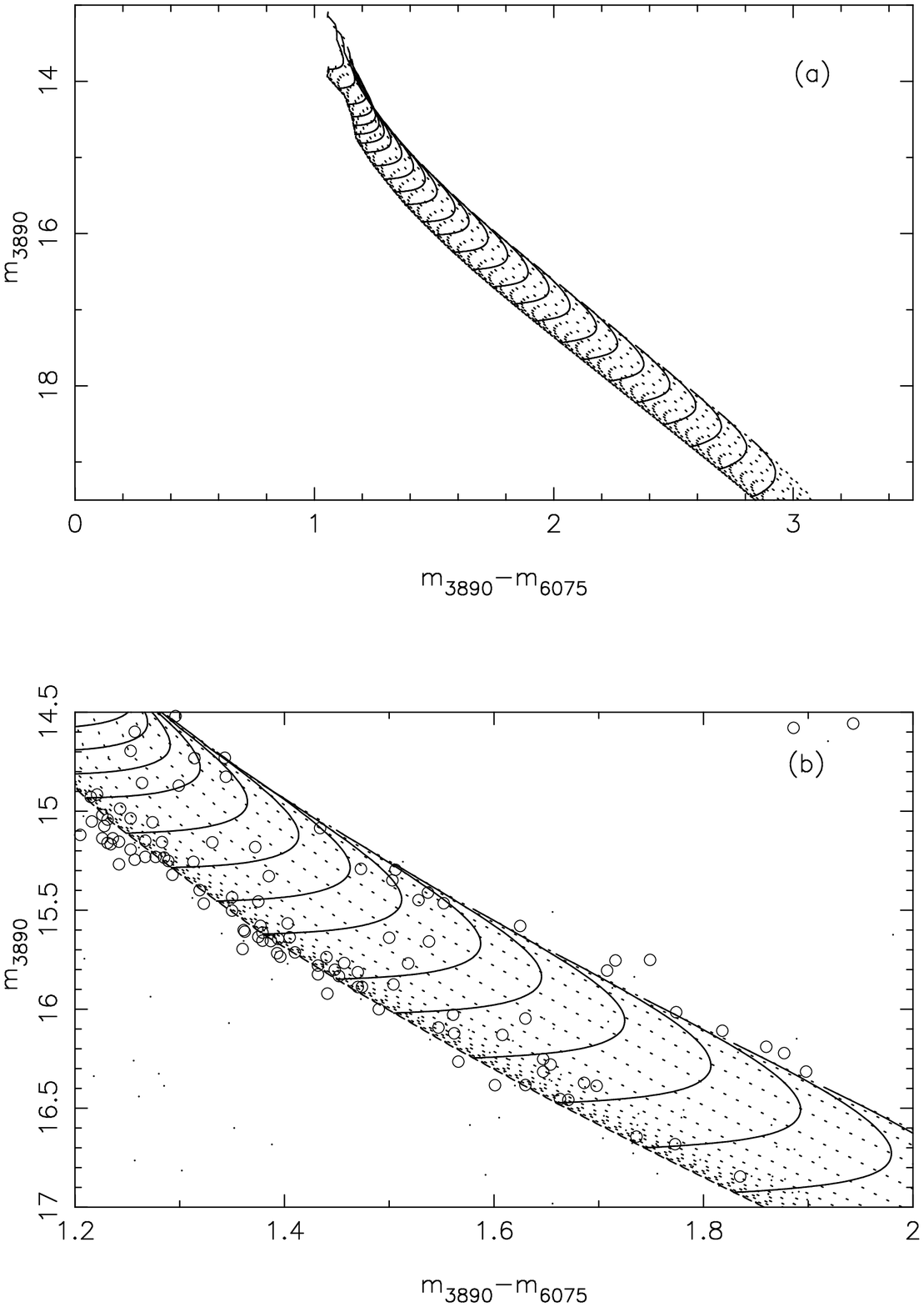}

\vspace{1cm}
\hspace{12cm}Fig 12. Fan et al.
\end{figure}
\begin{figure}
\epsfysize=600pt \epsfbox{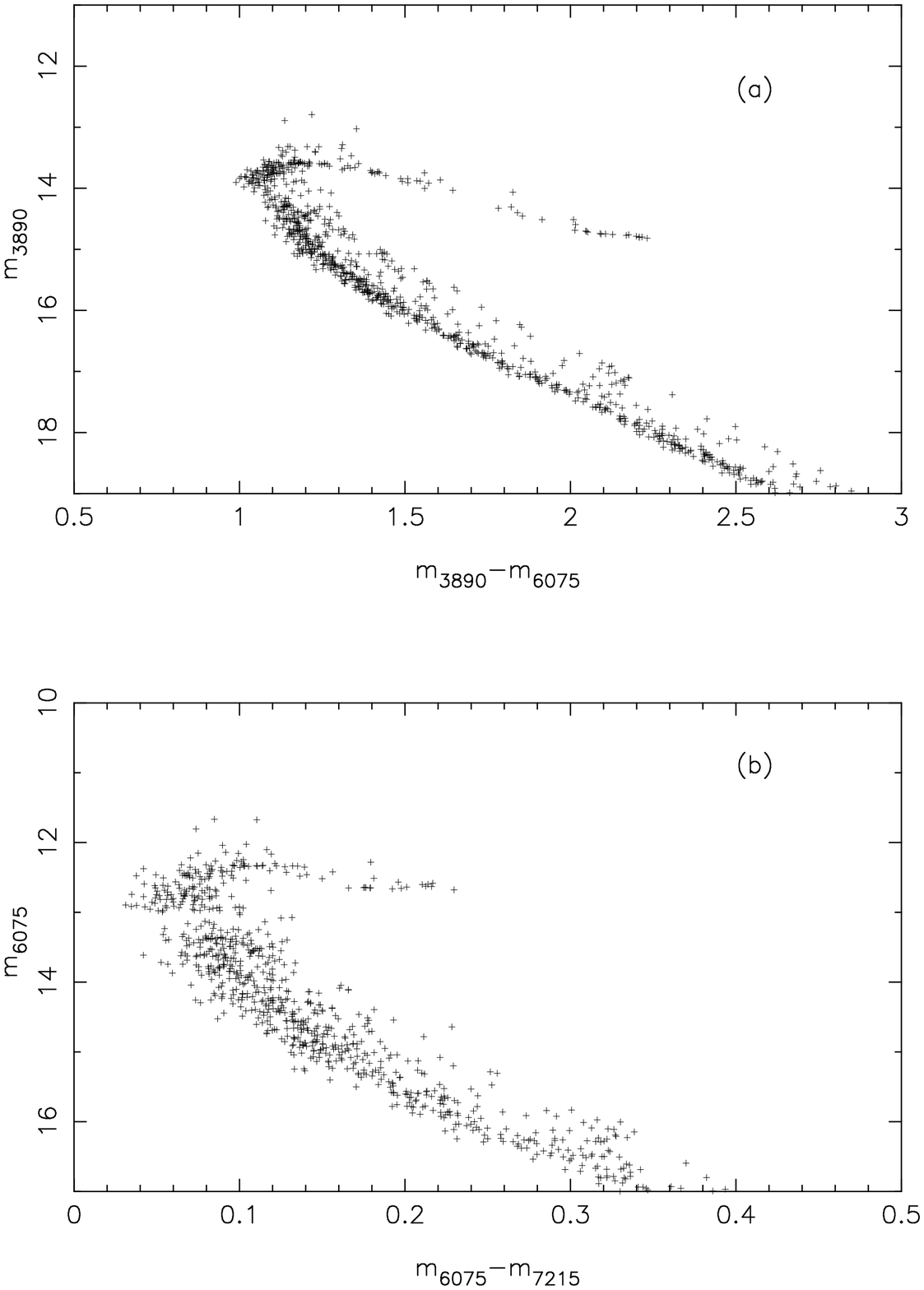}

\vspace{1cm}
\hspace{12cm}Fig 13. Fan et al.
\end{figure}
\begin{figure}
\vspace{-6cm}

\epsfysize=600pt \epsfbox{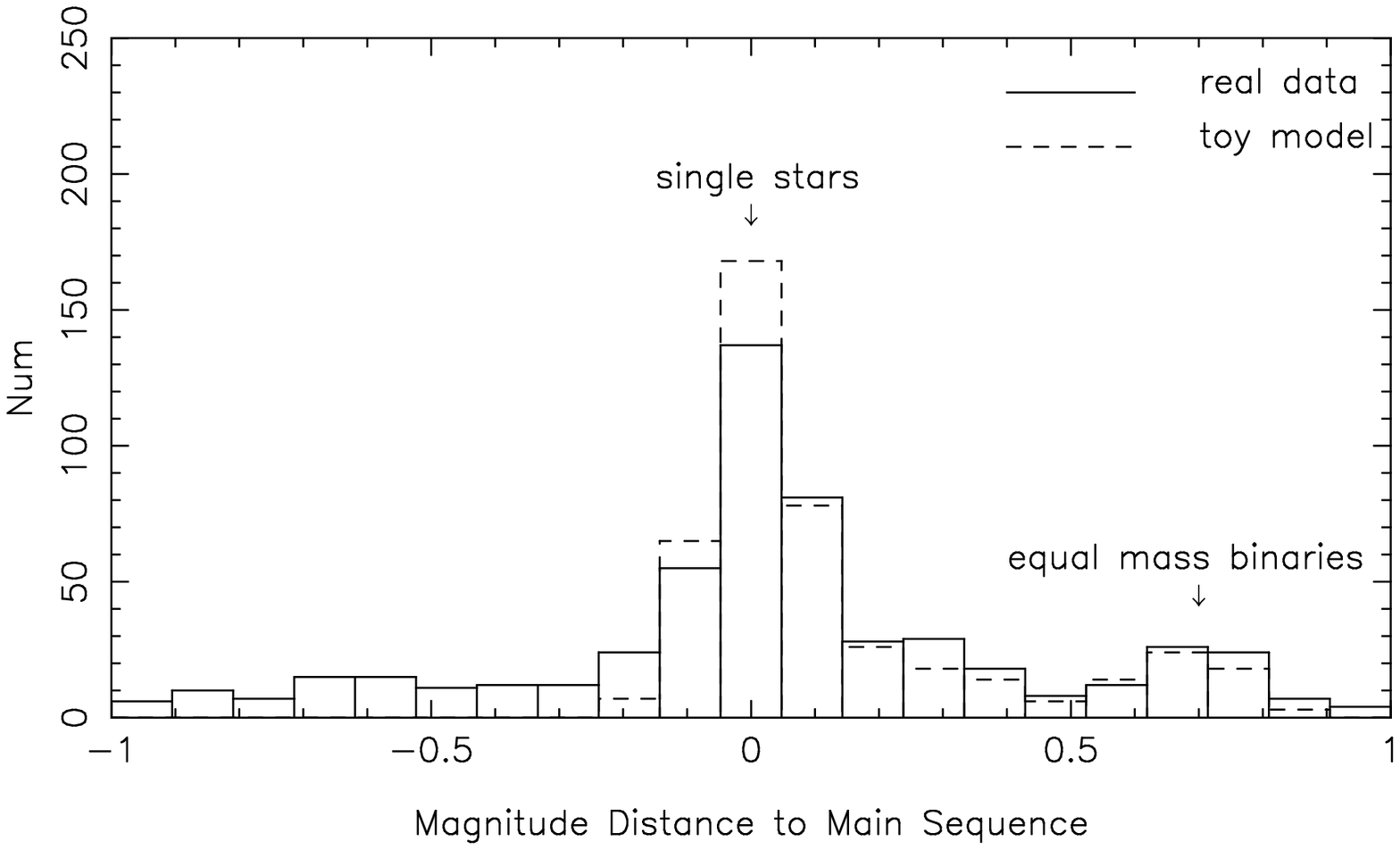}

\vspace{6.5cm}
\hspace{12cm}Fig 14. Fan et al.
\end{figure}
\begin{figure}
\vspace{-6cm}

\epsfysize=600pt \epsfbox{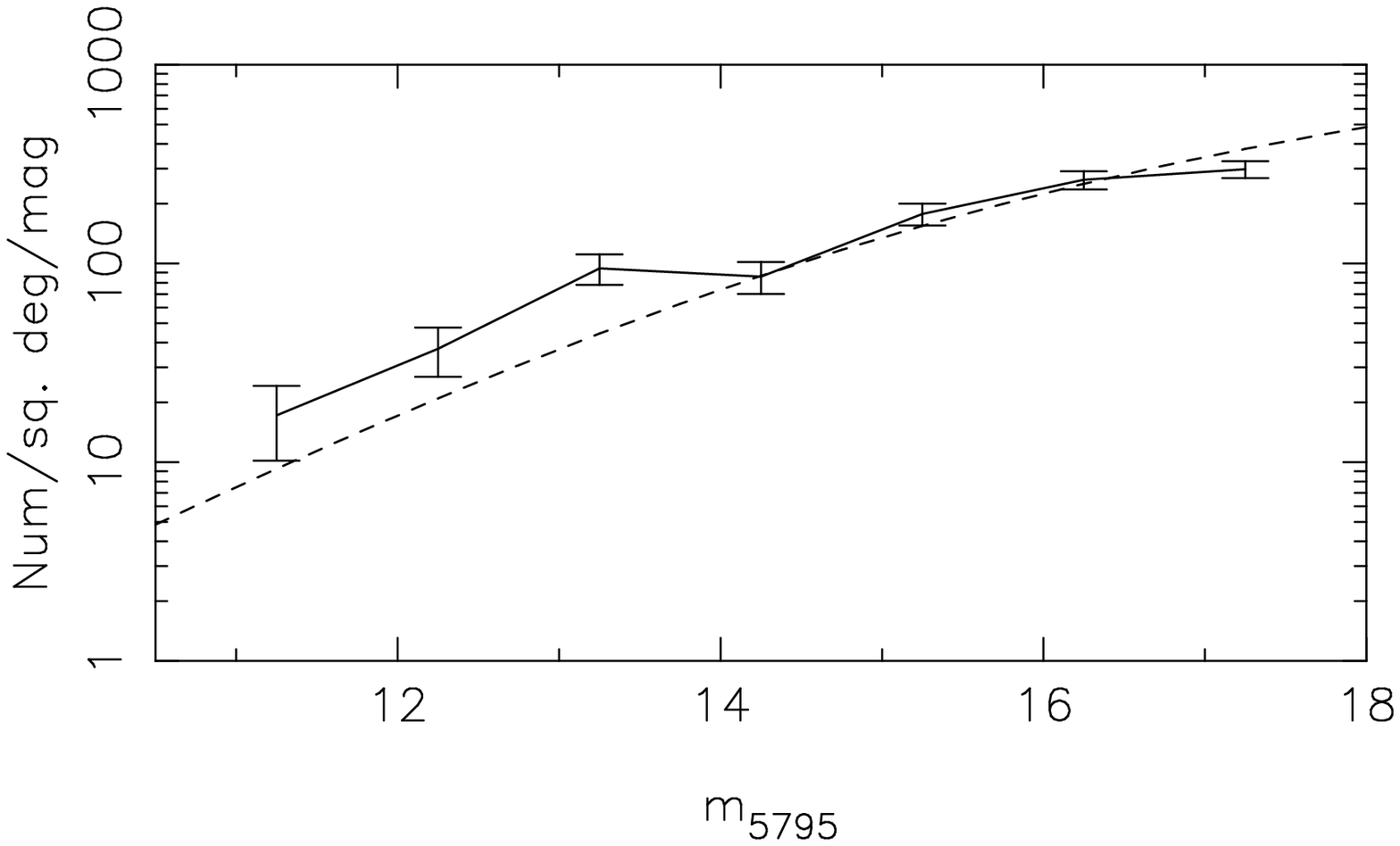}

\vspace{6.5cm}
\hspace{12cm}Fig 15. Fan et al.
\end{figure}
\begin{figure}
\epsfysize=600pt \epsfbox{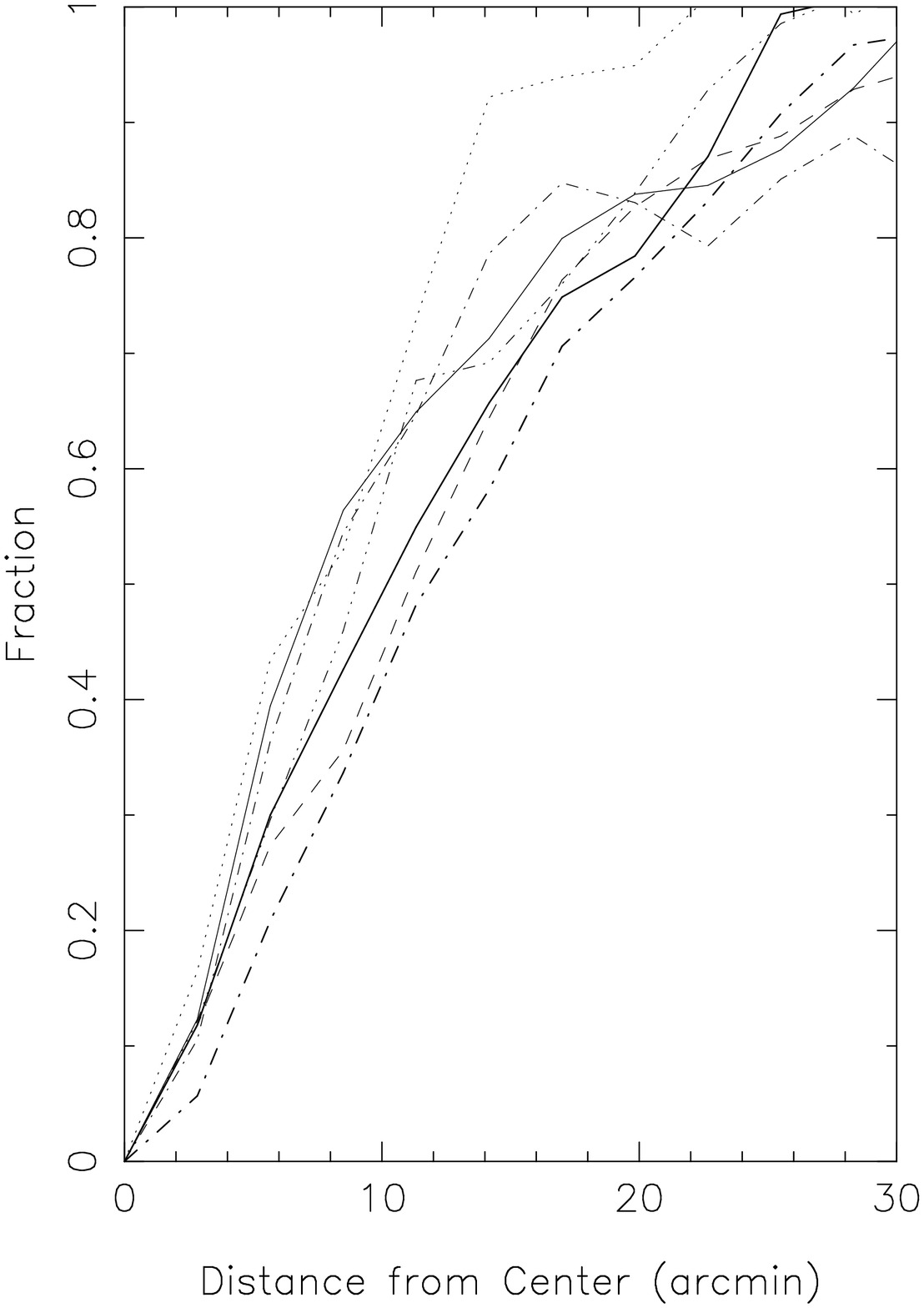}

\vspace{1cm}
\hspace{12cm}Fig 16. Fan et al.
\end{figure}
\begin{figure}
\epsfysize=600pt \epsfbox{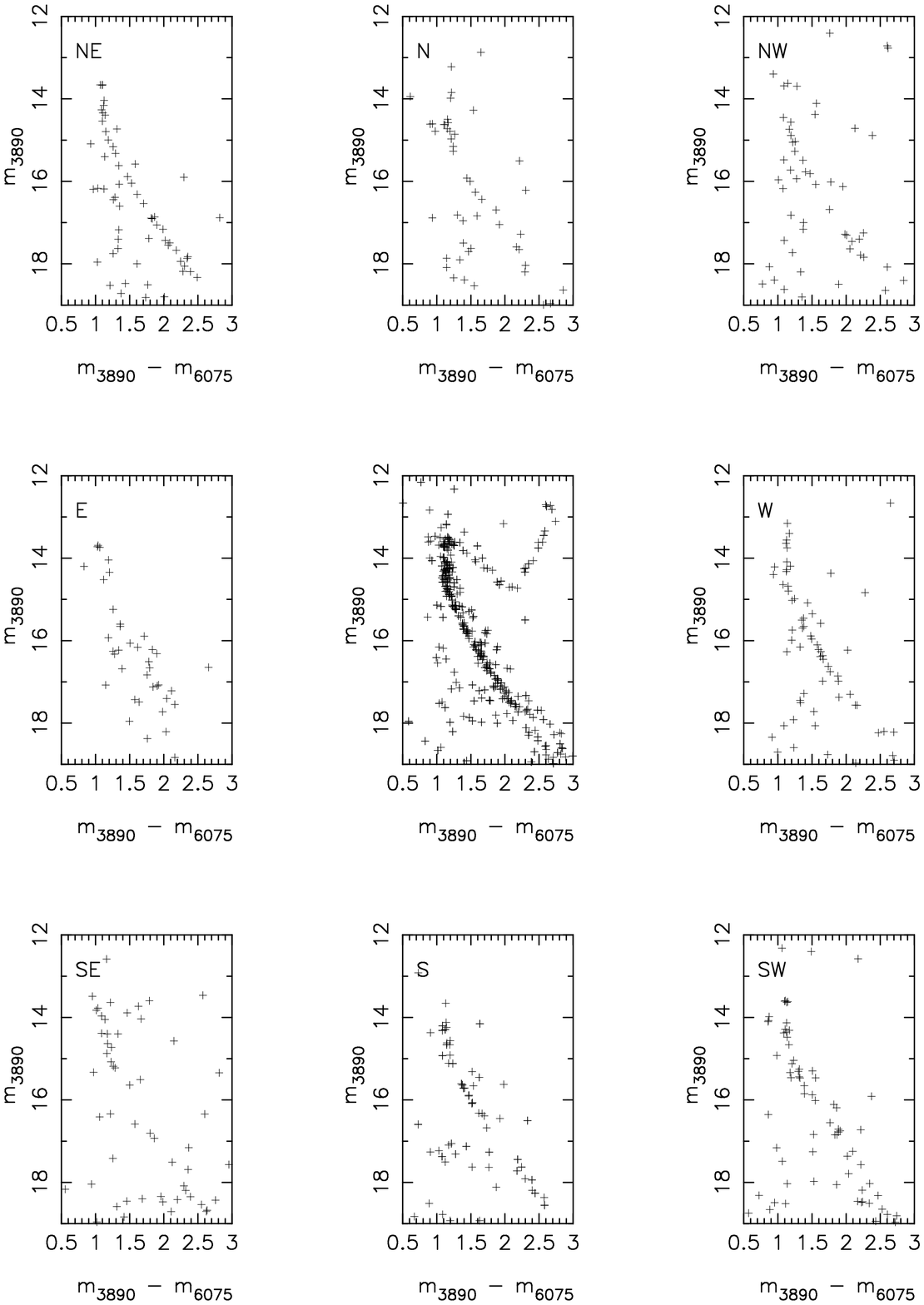}

\vspace{1cm}
\hspace{12cm}Fig 17. Fan et al.
\end{figure}
\newpage
\begin{figure}
\epsfysize=600pt \epsfbox{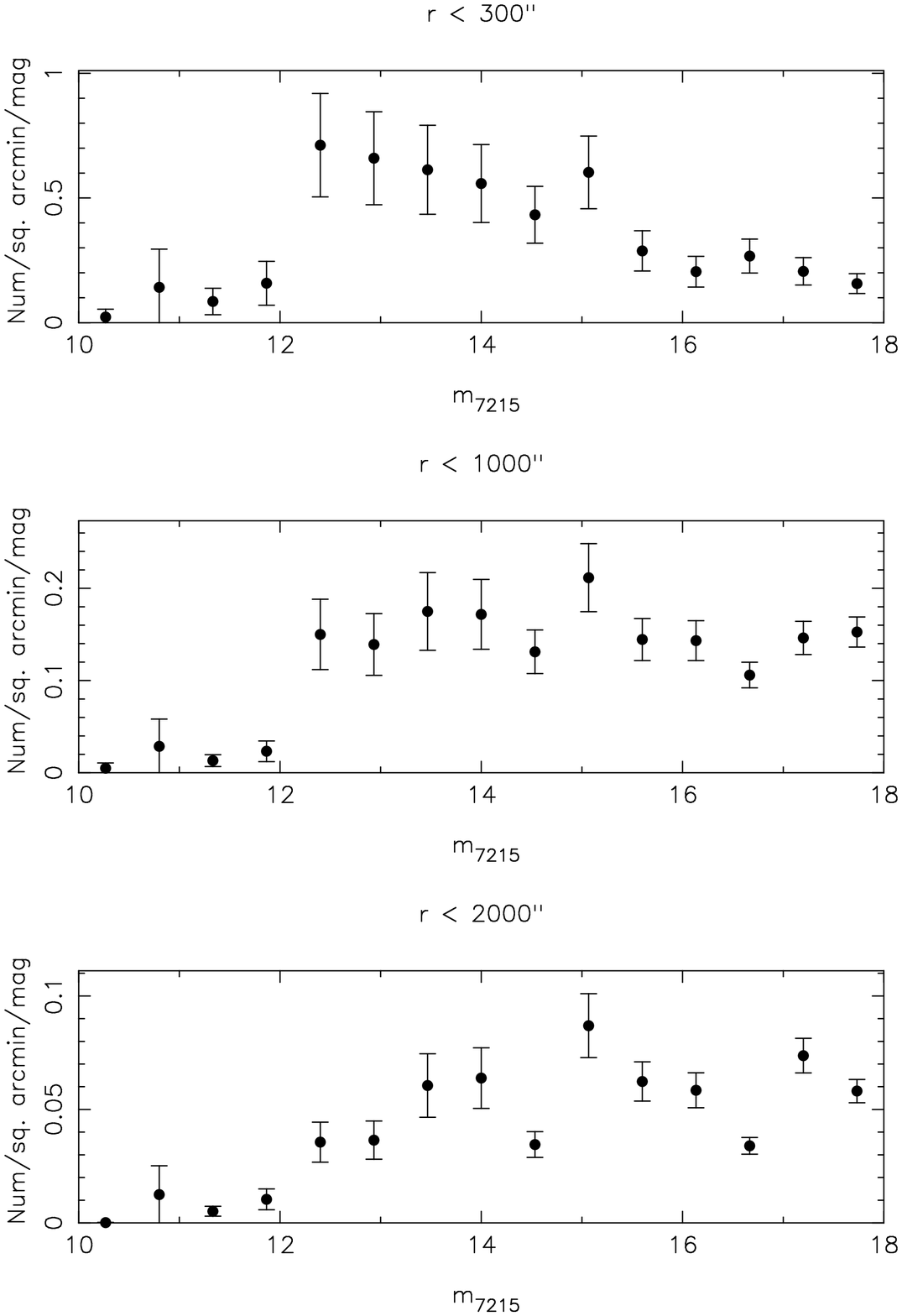}

\vspace{1cm}
\hspace{12cm}Fig 18. Fan et al.
\end{figure}
\newpage
\begin{figure}
\vspace{-6cm}

\epsfysize=600pt \epsfbox{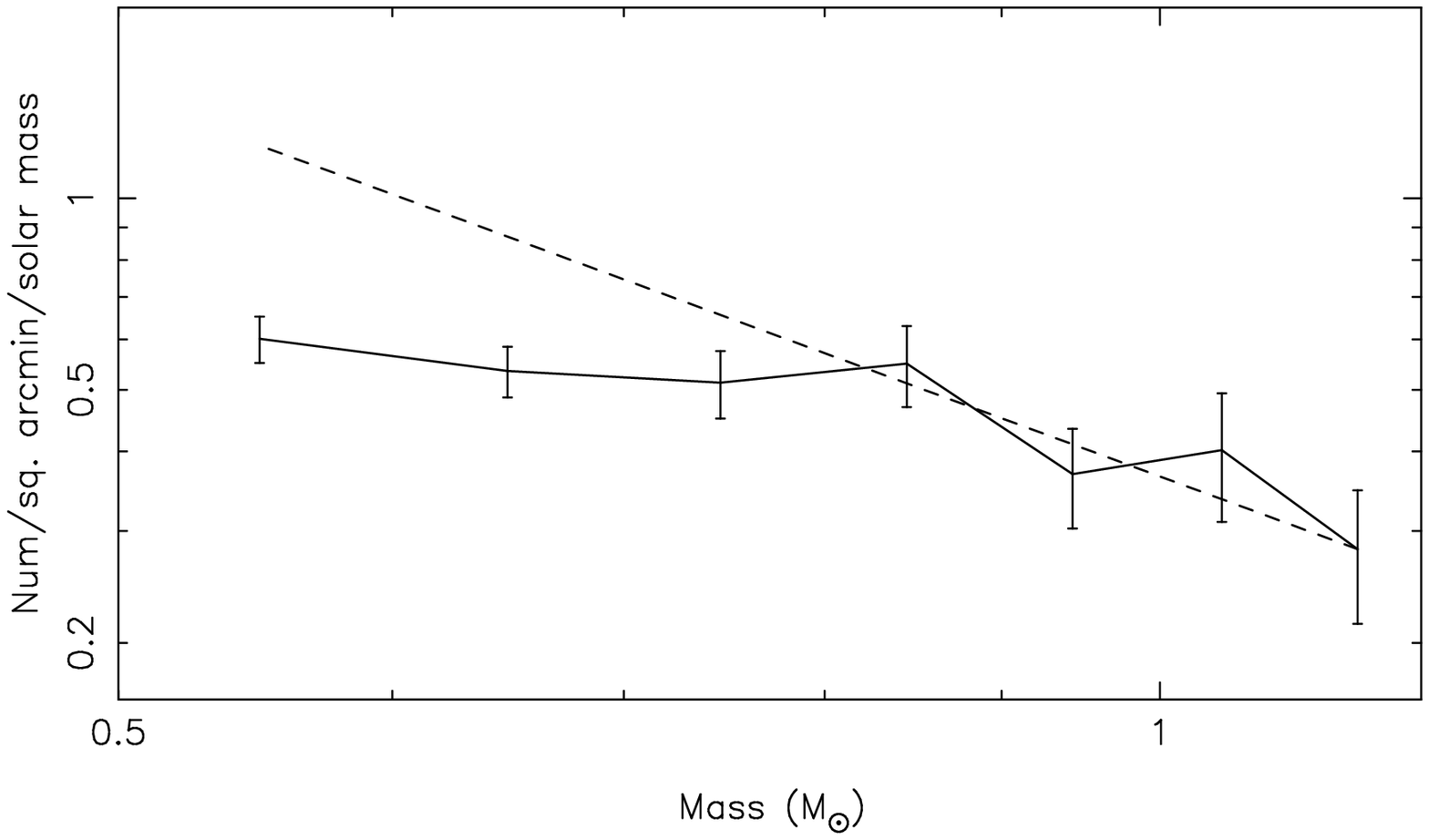}

\vspace{6.5cm}
\hspace{12cm}Fig 19. Fan et al.
\end{figure}
\end{document}